\documentclass[11pt]{article}
\pdfoutput=1

\usepackage{jheppub}
\usepackage{url,comment}
\usepackage{times}
\usepackage{latexsym}
\usepackage{graphicx, graphics, hyperref, amsmath, amssymb, slashed, color}
\usepackage{bbm,amsthm}
\newcommand{\lp}{\left(}
\newcommand{\rp}{\right)}
\newcommand{\order}[1]{\mathcal{O}\lp#1\rp}

\newcommand{\abs}[1]{\left| #1\right|}
\newcommand{\vev}[1]{\left\langle #1\right\rangle}
\newcommand{\nc}{\newcommand}

\nc{\beq}{\begin{equation}}
\nc{\eeq}{\end{equation}}
\newcommand{\beqa}{\begin{eqnarray}}
\newcommand{\eeqa}{\end{eqnarray}}
\newcommand{\bea}{\begin{eqnarray*}}
\newcommand{\eea}{\end{eqnarray*}}
\nc{\sq}{\tilde q}

\title{Stopping Quirks at the LHC}

\author[a]{Jared A.~Evans}
\author[b]{and Markus A.~Luty}
\affiliation[a]{Department of Physics, University of Cincinnati, Cincinnati, Ohio 45221, USA}
\affiliation[b]{Physics Department, University of California, Davis, Davis, California 95616, USA}

\emailAdd{jaredaevans@gmail.com}
\emailAdd{luty@physics.ucdavis.edu}

\abstract{
Quirks are exotic particles charged under a new confining gauge group that can 
give rise to unique collider signatures, 
depending on their vector-like mass, quantum numbers, and the confinement scale.   
In this work, we consider the possibility that quirks produced at the LHC lose all of 
their kinetic energy through ionization loss before escaping the detector,
and annihilate at a time when there are no active $pp$ collisions.  
We recast an existing CMS search for out-of-time decays 
of $R$-hadrons to place new limits on quirk parameter space.  
We propose several simple
modifications to the existing out-of-time search strategy that 
can give these searches sensitivity
in regions of quirk parameter space 
not covered
by any existing or proposed search strategy.}

\arxivnumber{nnnn.nnnnn}

\begin{document}

\maketitle

%%%%%%%%%%%%%%%%%%%%%%%%%%%%%%%%%%%%%%%%%%%%%%%%%%
\section{Introduction\label{sec:intro}}
%%%%%%%%%%%%%%%%%%%%%%%%%%%%%%%%%%%%%%%%%%%%%%%%%%

The large hadron collider (LHC) has set impressive limits on  
many different
models, but thus far has discovered no unambiguous signs for new physics beyond the
Standard Model. 
One place where new physics could be hiding is in exotic detector objects, such as 
long-lived particles,
which can arise in a variety of models, {\it e.g.}~\cite{Giudice:1998bp,Barbier:2004ez,Kang:2008ea,Arvanitaki:2012ps,Cui:2012jh}.
Long-lived particles and other exotic states will generally
evade detection unless searches are designed to look 
for the specific signatures characteristic of these particles 
\cite{Cui:2014twa,Liu:2015bma,Csaki:2015uza,Evans:2016zau,Coccaro:2016lnz,Liu:2018wte,Curtin:2018mvb,Lee:2018pag}.   
Such dedicated searches at CMS, ATLAS, and LHCb
are beginning to probe these
regions of parameter space as well 
(see {\it e.g.}~\cite{Khachatryan:2014mea,Aaboud:2016uth,Aaij:2016isa,Khachatryan:2016sfv,Aaij:2016xmb,Aaij:2017mic,Aaboud:2017iio,Sirunyan:2017jdo,Aaboud:2017mpt}).

Quirks are one type of 
long-lived particle
that has proven particularly difficult to probe
with the existing LHC search program \cite{Kang:2008ea}.  
Quirks are exotic 
fermions $Q$ charged under a new  
confining $SU(N)$ gauge group, 
referred to here as infracolor (IC).    
The lightest quirk has a vector like mass $m_Q$ that is assumed to be  
much larger than the confinement scale $\Lambda_{\text{IC}}$.   
This hierarchy of scales
implies that when the constituents of a $Q\bar Q$ pair are separated, the quirks remain connected by a macroscopic gauge flux tube 
\cite{Mandelstam:1974vf,Nambu:1978bd,Sundrum:1997qt}, 
referred to as the string, which imparts a 
constant force on the quirks, 
$|\vec{F}| \sim \Lambda_{\text{IC}}^2$.  
Since the local energy density stored in the string 
is set by $\Lambda_{\text{IC}}$, it
is never enough to pull new $Q\bar Q$ pairs out of the vacuum, 
and therefore the string will not break.  
If the quirks additionally have standard model charges,  the quirks on the end of the string can be produced at the LHC and interact with the detector.  Quirks that are produced have a characteristic length scale for the string,
\beq
\ell \sim \frac{m_Q}{\Lambda_{\text{IC}}^2} \sim 10\mbox{ cm} 
\lp \frac{m_Q}{\text{1 TeV}}\rp 
\lp\frac{\text{1~keV}}{\Lambda_{\text{IC}}}\rp^2,
\label{eq:length}
\eeq
which allows for 
quirks with different values of $m_Q$ and $\Lambda_\text{IC}$
to give rise to a wide variety of
possible collider signatures \cite{Kang:2008ea}.  

For length scales less than the detector resolution ($\ell\lesssim100~\mu$m),
the signature is a straight, highly-ionizing track, 
which has
been searched for at D0 \cite{Abazov:2010yb}.  
Over a wide range of confinement scales, mono-jet searches 
\cite{CMS:2016pod,Aaboud:2016tnv} can constrain quirks 
because the sparse collection of hits that quirks 
leave 
in the detector are not typically reconstructed by the tracking algorithm 
\cite{Farina:2017cts}.  
For length scales larger than the detector size ($\ell\gtrsim10~$m), where the effect of the string force is minimal, heavy stable charged particle (HSCP) searches \cite{CMS:2016ybj} are able to place stringent limits on quirks \cite{Farina:2017cts}.   
A proposal to look in the monojet sample for coplanar hits that are unmatched to tracks may
cover intermediate length scales (1 mm $\lesssim\ell\lesssim 1$ m) \cite{Knapen:2017kly}.  These efforts are only the first steps made in accessing the variety of novel quirky signatures that could potentially be 
discovered at colliders.

Another possible search strategy is to target those
quirks that stop within the 
experiment  
due to interactions with the detector 
material, then 
 annihilate out-of-time with the active $pp$ collisions.
These produce the 
novel signature 
 of energy deposits in the calorimeters when no collisions are 
taking place.
Searches for stopped particles or out-of-time decays already exist
at both ATLAS \cite{Aad:2013gva,Aad:2012zn} and CMS \cite{Sirunyan:2017sbs,Khachatryan:2015jha,Chatrchyan:2012dxa}.  
These searches were designed to search for $R$-hadrons,
long-lived colored supersymmetric particles, such as the gluino or stop,
that 
hadronize, 
lose their kinetic energy through material interactions,
and then decay long after production \cite{Arvanitaki:2005nq}.
These out-of-time searches are always less sensitive than searches 
for $R$-hadron tracks 
(HSCP searches), 
although the out-of-time approach could provide valuable information about 
the lifetime of a heavy superpartner discovered in a HSCP search.

However, the out-of-time annihilation of quirks exhibits substantially  
different properties than the out-of-time decay of an $R$-hadron.  
In particular, quirks are typically produced more forward and annihilate into
significantly more visible energy than an $R$-hadron of the same mass would.  
We will show that simple modifications
to the existing search efforts that capitalize on these features 
can greatly enhance sensitivity to quirks.  
In particular, 
the kinematics of quirks make them a perfect target signature for the new CMS endcap calorimeter that will be installed for Run IV \cite{Collaboration:2293646}. 

In this paper, we recast the 13 TeV search at CMS for the
out-of-time decay of stopped particles to place new limits on quirks. 
This paper is organized as follows.  
In section~\ref{sec:stopping}, we will discuss the production of quirks and their energy loss in the CMS 
detector, with simulation details discussed in appendix~\ref{sec:CMS}.
The resulting out-of-time annihilation and trigger efficiency 
is discussed in section~\ref{sec:oot}.  
The results of recasting existing searches are presented in section~\ref{sec:limits}.
In section~\ref{sec:improve}, we discuss improvements to the out-of-time search strategy
that can greatly enhance sensitivity to quirks across a broad range of confinement scales. 
In appendix~\ref{sec:Eloss}, we discuss energy loss mechanisms that are relevant
for quirk pairs for short string lengths where ionization energy loss no longer
applies.
 
 %%%%%%%%%%%%%%%%%%%%%%%%%%%%%%%%%%%%%%%%%%%%%%%%%% 
\section{Quirk Motion within the Detector\label{sec:stopping}} 
%%%%%%%%%%%%%%%%%%%%%%%%%%%%%%%%%%%%%%%%%%%%%%%%%%

Quirks with Standard Model
quantum numbers can be pair produced in LHC collisions through QCD or Drell-Yan-like interactions.  In this work, we will focus on two different cases of quirk standard model gauge quantum numbers: quirks charged as a right-handed charged lepton and as a right-handed up-type quark.  In 
(SU$(N)_{\text{IC}}$,\,SU$(3)_{C}$,\,SU$(2)_{L})_{U(1)_Y}$ 
notation, these are
\beq
E' \sim (N, 1,1)_{1}, \hspace{8mm} T' \sim (N, \bar 3,1)_{-\frac{2}{3}},
\eeq
together with the corresponding vector-like anti-particle.  
We will consider the specific case where the infracolor 
gauge group is $SU(2)$, since
this will lead to the most conservative constraints.    
Larger groups lead to larger production
cross-sections ($\sigma \propto N$), and lower branching fractions into invisible infracolor glueballs (see section~\ref{sec:limits} for more details).   
In both the $E'$ and $T'$ models, a pair of quirks with opposite electric charge
is produced, and these propagate through the detector under the influence of the
infracolor string force.
In the $T'$ model, the quirks also undergo hadronization due to QCD color interactions,
forming hadrons that could be charged or neutral at the ends of the string.%
\footnote{In less than 1\% of events, a baryon with charge two, such as $T'uu$, can form.  While these particles are included and modeled in our simulation, they are rare enough that they make no qualitative impact on  the results.} 
To handle these effects, we modified
Pythia 8 \cite{Sjostrand:2014zea} to allow for a wide range of quirky hadrons to be 
formed.
For simplicity all possible quirky hadrons are assumed to be detector stable, which is 
supported by spectrum expectations from HQET \cite{Neubert:1993mb}.

    \begin{figure}[t]
\begin{center}
\includegraphics[scale=1.]{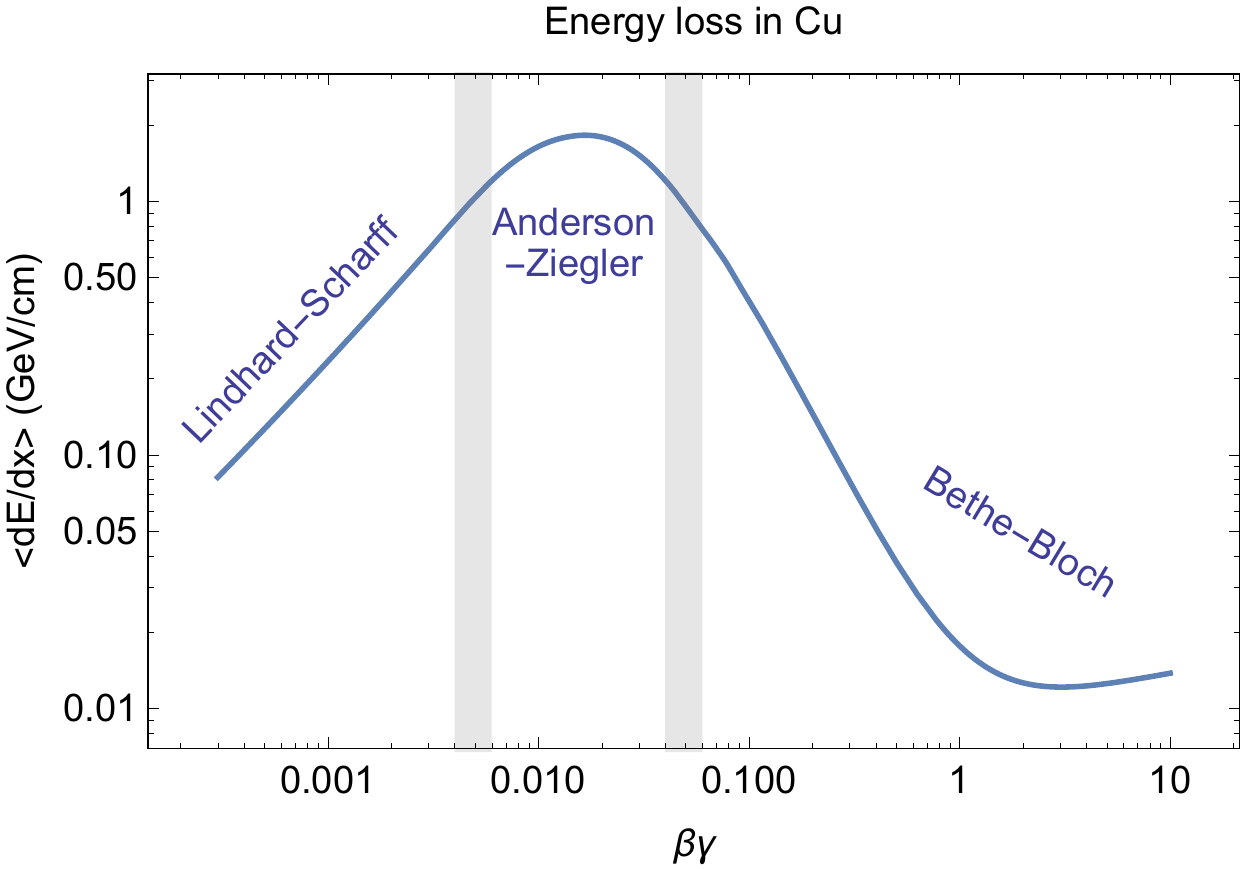}
\end{center}
\caption{ The average ionization loss of a charged particle moving through solid copper as a function of $\beta\gamma=v/\sqrt{1-v^2}$.   The Bethe-Bloch and Lindhard-Scharff regions are well predicted by theory \cite{Patrignani:2016xqp}.  The Anderson-Ziegler region is an interpolation from data, and the gray bands denote the interface between these regions.}  
\label{fig:BB}
\end{figure}

The quirk pairs are produced in a bound state with a highly
excited radial mode and low angular momentum. 
In the lab frame, we can view the quirk system as having three separate components to the total energy, 
 \beq
 E_{\text{tot}} = 2 m_Q + E_{\text{int}} + E_{\text{cm}},
 \label{eq:Edist}
 \eeq
  the rest mass of the two quirks, the internal energy, and the center of mass energy.  The invariant mass of the quirk pair system is $m_{\text{int}} = 2 m_Q + E_{\text{int}}$.   The internal energy $E_{\text{int}}$ begins as kinetic energy, but it is turned into potential energy stored in the string as the quirks move apart.  As the quirks oscillate, this internal energy moves from kinetic energy of the particles to potential energy in the string and back again.  The center-of-mass energy $E_{\text{cm}}$ is the kinetic energy contributing to the bound state's motion through the detector, 
{\it i.e.}~in the quirk pair rest frame the center of mass energy is zero.

A charged particle moving through matter will lose energy via Bethe-Bloch ionization loss \cite{Patrignani:2016xqp}, see Fig.~\ref{fig:BB}.   
Since 
we are primarily concerned with particles losing all of their energy rather than the energy deposited in individual detector systems, it is reasonable to consider the mean energy loss $\vev{\frac{dE}{dx}}$ rather than the most probable energy loss $\Delta_p$. 
We also use a continuous
friction approximation rather than sampling the energy loss from a Landau distribution.  These approximations make the propagation code computationally efficient,
and are not expected to 
change the qualitative results.
The specifics of our CMS detector simulation are described in appendix~\ref{sec:CMS}.

For colored quirks, nuclear interactions can provide an additional source of energy loss, 
and allow for charge flipping to occur within the dense calorimeters.  
Generically, the inclusion of these effects ultimately results in a larger stopping fraction 
than the case of ionization loss alone \cite{Aad:2013gva}.
For this study, we conservatively neglect nuclear interactions.   
QCD hadronization takes place independently for each quirk in a pair, and
can result in either electrically charged or neutral quirky hadrons
connected by an infracolor string.
In this approximation neutral quirky hadrons
do not lose energy in the detector.
  
The quirk motion is additionally affected by both the infracolor force 
due to the string  and the magnetic field in the detector.
While no energy is directly lost through these mechanisms, the modification to the quirk motion through the material greatly alters how much energy is lost overall.   The infracolor force is \cite{Kang:2008ea}
  \beq
 \vec F_{\text{IC}}  = - \Lambda_{\text{IC}}^2 \lp \sqrt{1-{v}_\perp^2} \hat{\text{s}}+ \frac{\abs{v_\|}}{\sqrt{1-{v}_\perp^2}}  \vec{v}_\perp \rp.
\eeq
where $\hat{\text{s}}$ is the direction of the string connecting to the quirk, and $\vec{v}_\|$ and $\vec{v}_\perp$ are the velocity of the quirks defined with respect to $\hat{\text{s}}$.    In simulating the quirk propagation, we make the simplifying ``straight-string approximation'' which assumes that in the quirk pair rest frame the string is straight and points between the two quirks so that $\hat{\text{s}} \propto (\vec x_1-\vec x_2)$.  As the quirks begin with no relative angular momentum and typically acquire very little relative to the overall excitation of the system, retarded time corrections can also be safely neglected.  The net equation of motion governing the propagation of each quirk is then
\beq
\frac{d\vec{p_i}}{dt} = \vec F_{\text{IC}}   - q_i^2 
\left\langle \!\frac{dE}{dx}\!\right\rangle\!(\vec x_i, v_i)\, \hat v_i 
+  q_i \vec v_i \times \vec B(\vec x_i)
\eeq
where $q_i$ is the charge of the quirk or quirky hadron, and the strength of the magnetic field in the barrel and end caps is approximately matched to data \cite{Chatrchyan:2009si}.

    \begin{figure}[t]
\begin{center}
\includegraphics[scale=1.]{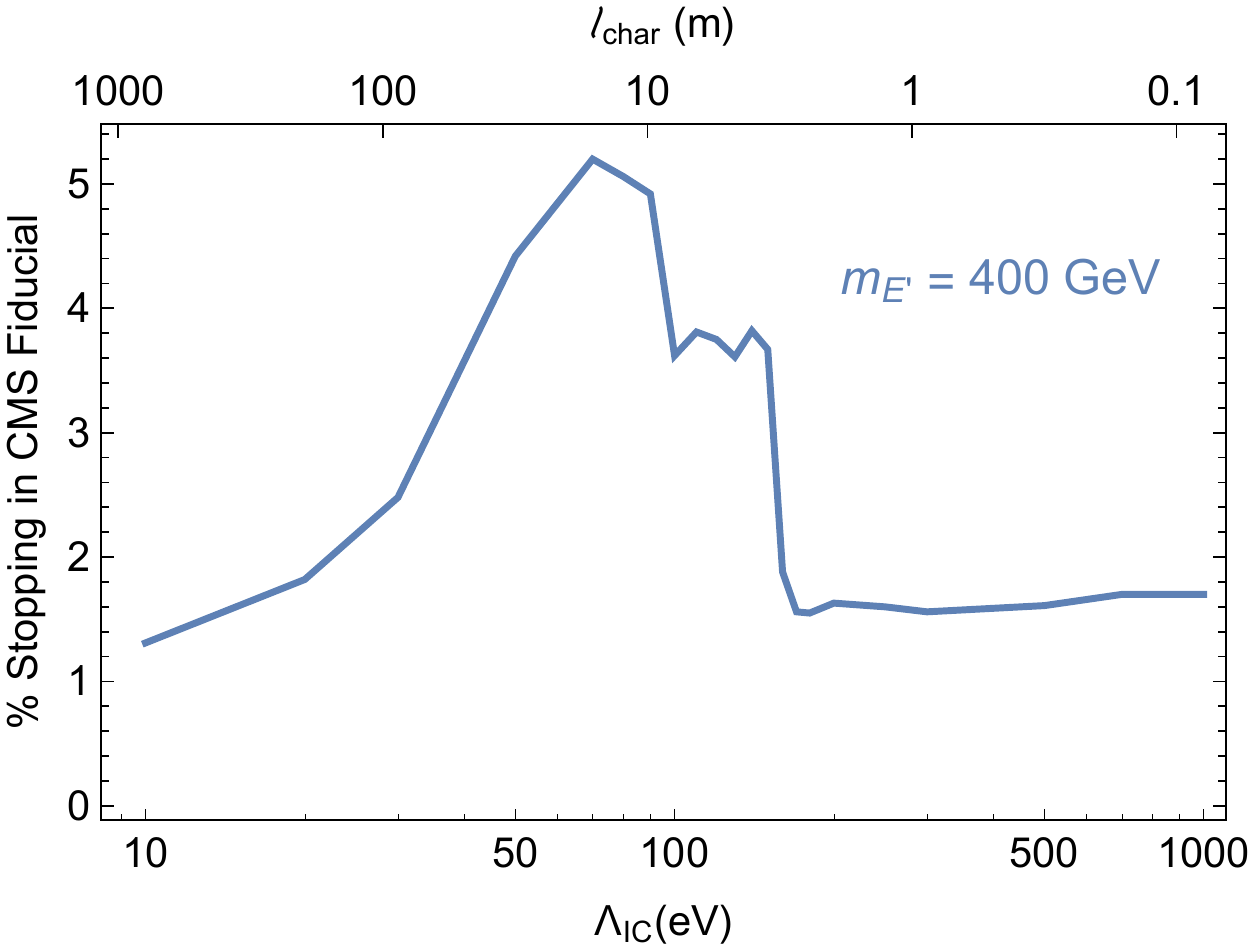}
\end{center}
\caption{Percentage of $E'$-like quirk pairs that stop within the CMS fiducial region (ECal and HCal with $\abs\eta<1.0$) for $m_{E'}=400$ GeV.   Above $\Lambda_{\text{IC}}\sim 100$ $(150)$~eV, the quirks no longer have a terminal velocity in the ECal (HCal), reducing the likelihood to stop.  The characteristic separation length Eq.~(\ref{eq:length}) is shown above.}  
\label{fig:Fiduc}
\end{figure}

 In the case of low confinement scales ($\Lambda_{\text{IC}}\lesssim 150$~eV),  the $\left\langle \frac{dE}{dx}\right\rangle$ value in the Anderson-Ziegler peak (Fig.~\ref{fig:BB}) can be greater than $\abs{F_{\text{IC}}}$ 
 (1 GeV/cm $=$ (140 eV)$^2$), 
 so that once the quirk velocity falls below this peak, the quirks have a terminal velocity $\beta_{\text{max}}\lesssim 10^{-2}$.     For larger confinement scales, there is no terminal velocity and the quirks will oscillate freely.  For these large confinement scales, the characteristic length is short enough that the quirks will typically undergo many oscillations during their motion
through the detector.  These oscillatory trajectories result in the quirks moving through more material over a much longer period of time than would be observed with a typical heavy stable charge particle.  This effect is amplified when $E_{\text{int}}\gg E_{\text{cm}}$.   
Also, because
charged particles lose more energy when they move slowly (see Fig.~\ref{fig:BB}), and quirks tend to be moving slowest when the internal energy is stored entirely in the string, energy is lost more quickly from the center-of-mass kinetic energy than from the internal energy Eq.~(\ref{eq:Edist}).  As a consequence, the quirks will often lose their center-of-mass energy while still having a fairly large internal energy remaining, so that they oscillate back-and-forth within the detector before annihilating.   All of these effects work to increase the likelihood that the quirk pair will stop within the detector and annihilate out-of-time.\footnote{One potential effect not considered here is whether positively charged quirks can reliably strip electrons and become a neutral state.  This effect is small for other particles, but quirks can briefly assume a low velocity while still having a lot of energy to lose, which may increase this likelihood.    On the other hand, the infracolor force itself may eject bound electrons.}
 
In Fig.~\ref{fig:Fiduc}, we show the percentage of the $E'$-like quirk pairs produced 
in 13 TeV collisions that stop within the CMS fiducial volume, which includes 
both the ECal and HCal detector systems for pseudorapidity $\abs\eta<1.0$.  
At $\Lambda_{\text{IC}}\sim 10$~eV, the characteristic separation length is of order
1~km, so the infracolor force does very little to impede the quirks, and they effectively
need to stop independently within the detector before slowly drifting toward one another at
their very low terminal velocity.\footnote{At confinement scales even lower than this, the quirks can bind with the lattice and may never annihilate \cite{Kang:2008ea}.}   
As the confinement scale increases, the infracolor force is more effective at slowing the
quirks, and the characteristic length becomes of order the detector size, so the probability
for a pair to stop rises.    
Above $\Lambda_{\text{IC}}\sim 100$~eV, the quirks no longer achieve a terminal velocity
within the ECal.   
Above $\Lambda_{\text{IC}}\sim 150$~eV, the quirks no longer have a terminal velocity in
the HCal.  
The efficiency briefly drops as the dense iron return yoke of the muon system continues to
provide a terminal velocity, slightly reducing the likelihood that quirks will stop in the
calorimeters.  
At large confinement scales, $\Lambda_{\text{IC}}\gtrsim 300$~eV, the probability for the
quirks to stop does not change appreciably with increasing confinement scale.  
This is because the energy loss over one oscillation has a minimal impact on the quirk
trajectory and each quirk travels approximately
the same amount of distance independent of the confinement scale.%
\footnote{Due to the computational feasibility of modeling the propagation when oscillation length gets very small, we only model confinement scales up to $\Lambda_{\text{IC}}=1$~keV.  We verified up to $\Lambda_{\text{IC}}\sim10$~keV that the stopping location and time is not appreciably perturbed from the 1~keV value.}    
The details of where these transition regions occur is in part a product of our CMS simulacrum, a more realistic treatment would surely induce small shifts in the precise value of these peaks and dips, but the qualitative features are expected to remain intact.
 
  %%%%%%%%%%%%%%%%%%%%%%%%%%%%%%%%%%%%%%%%%%%%%%%%%%
 \section{Out-of-Time Annihilation\label{sec:oot}}
 %%%%%%%%%%%%%%%%%%%%%%%%%%%%%%%%%%%%%%%%%%%%%%%%%%

Out-of-time decay searches capitalize on the timing (bunch train)
structure of the LHC beams in order to achieve low backgrounds.  
The LHC ring at a given time consists of 3564 distinct 25~ns bunch crossing windows \cite{Bailey:691782,FillSchemes1,FillSchemes2}.  In a typical fill of the ring, roughly two thirds of the bunch windows are occupied with colliding bunches, one third are empty, and a few contain collisionless bunches.  The CMS search looks within empty bunch windows that are at least two bunch crossings away from any colliding or non-colliding bunches.

\begin{figure}[t]
\begin{center}
\includegraphics[scale=1]{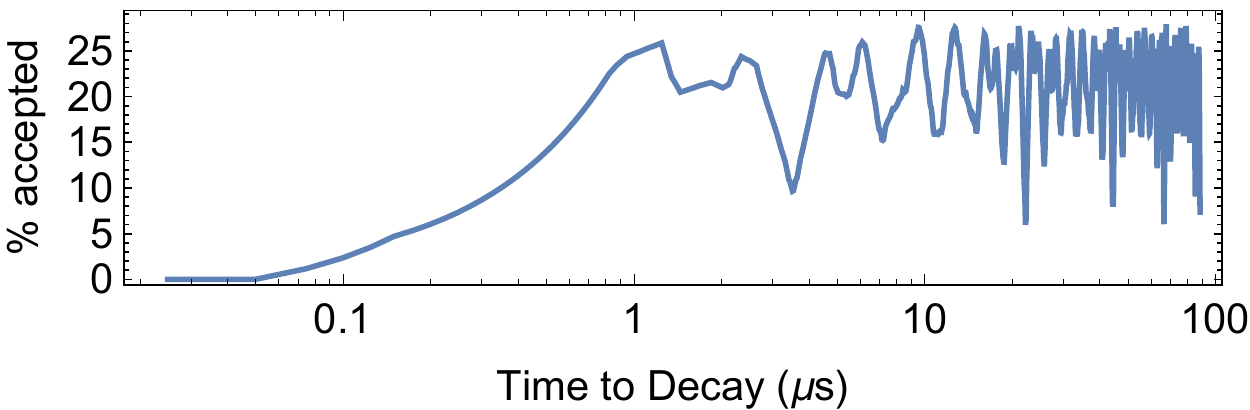}
\end{center}
\caption{The percentage of produced particles decaying at a later time $t-d/c$, where $d$ is the distance to the point of decay, that decay within a triggerable bunch window (at least two bunches away from filled bunch windows).  For decay times $>1 \mu$s, the average is 21\%.}  
\label{fig:decaytime}
\end{figure}

For 2015--2016 data, most of the colliding bunches were chained together within a densely packed ``batch'' of 
$\sim 50$
sequential bunches with no empty bunches in between them (see \cite{Bailey:691782,FillSchemes1,FillSchemes2} for more details).  Between each batch, there is a gap of roughly 8 empty bunch crossings.  After typically 2 to 4 batches, there is a larger injection gap of 
$\sim 40$
empty bunch crossings.    Lastly, there is a single large abort gap in the beam of 
$\sim 120$
empty bunch crossings.   
The specifics of the bunch train filling scheme varies frequently during LHC running \cite{FillSchemes2}.  
For the purposes of this study, we consider the exact pattern for the three largest fill schemes during 2016 that make up most of the data.
We determine as a function of true displacement the probability for a quirk pair 
annihilating at time $t-d/c$ to annihilate within an empty bunch window.
The results are shown in Fig.~\ref{fig:decaytime}.   
For decay times larger than 4~$\mu$s, we assume the average probability to decay 
within an acceptable trigger window, which is
21\%.  As the other fill schemes that are not used in this study typically are less densely packed, this value is conservative.

The annihilation of a quirk pair takes place in several stages.
After production, each quirk in the pair will first propagate through the detector 
subject to the string force, magnetic field, 
and frictional forces from ionization energy loss.
We consider events where these effects combine to stop the quirks and bring them to
rest inside the detector.
The quirk pair will then de-excite and annihilate.
The total time needed for this is therefore
\beq
\tau_{\text{tot}} = \tau_{\mbox{\scriptsize stop}} 
+ \tau_{\mbox{\scriptsize de-excite}} 
+ \tau_{\mbox{\scriptsize decay}}.
\label{eq:tau}
\eeq
For the purposes of this study we will approximate $\tau_{\text{tot}} \simeq \tau_{\mbox{\scriptsize stop}}$, and assume that all other processes are rapid.
The energy loss of a quirk pair becomes difficult to model when the size of the quirk
pair is smaller than atomic sizes.
In appendix~\ref{sec:Eloss}, we estimate the time to de-excite 
due to several mechanisms: electromagnetic radiation, infracolor interactions,
and electric currents induced in the material by the oscillating dipole.
These estimates support the assumption above for most of the parameter space.
We emphasize that the assumption that the de-excitation process is prompt is a 
conservative one, since increasing the time to annihilation will make the search
more sensitive, as long as the de-excitation time is shorter than $10^4$ seconds \cite{Sirunyan:2017sbs}, 
as is strongly suggested by multiple energy loss mechanisms considered in appendix~\ref{sec:Eloss}.

\begin{figure}[t]
\includegraphics[scale=0.58]{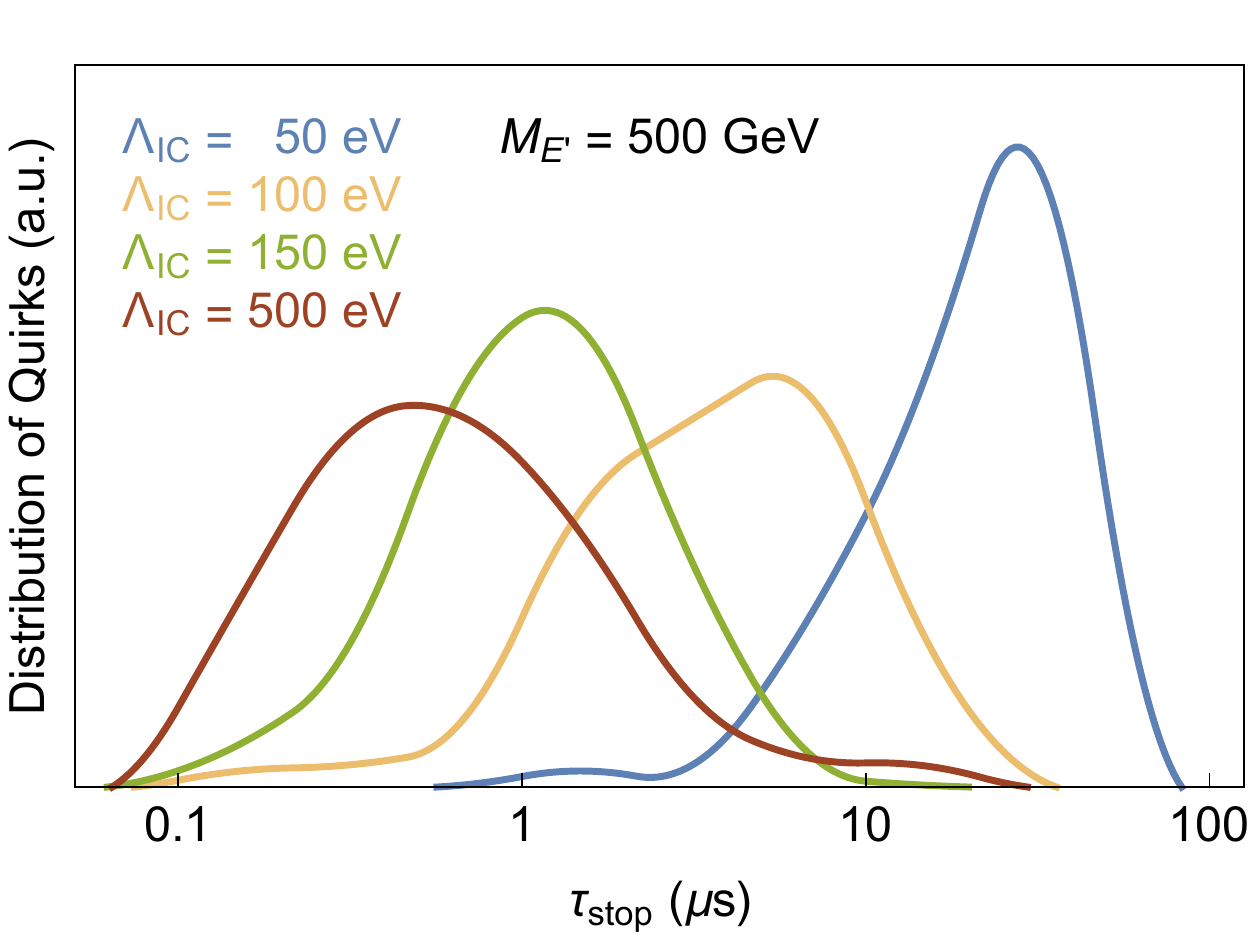}\vspace{-57mm}\\

\hfill \includegraphics[scale=0.63]{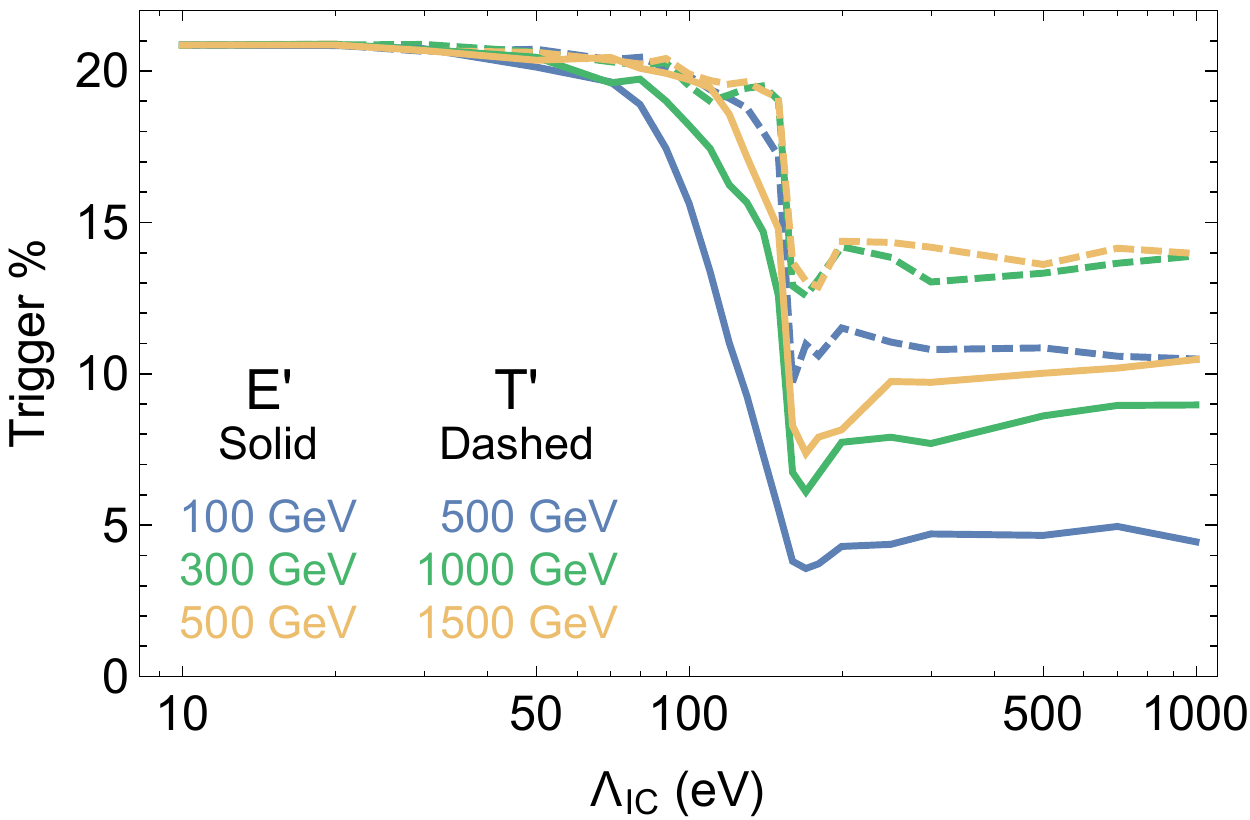}
\caption{{\bf Left:} 
Distribution of the time for $E'$-like quirks at 500 GeV to stop within the CMS fiducial region for several quirk confinement scales.  
For confinement scales above about 300~eV, the distribution does not change appreciably 
with increasing confinement scale.  
Decreasing the confinement scale once terminal velocities are achieved 
(see section \ref{sec:stopping}) decreases the terminal velocity greatly, 
thus increasing the time to stop.  
The time to annihilate after stopping is conservatively assumed to be much quicker. 
{\bf Right:}  Probability for quirks that stop in the CMS fiducial region to annihilate during a triggerable bunch window (at least two bunches away from filled bunch windows).  If the quirks can achieve a terminal velocity in the material, then the trigger percentage approaches the maximum of 21\%.   For $\Lambda_{\text{IC}} \gtrsim 300$~eV, the percent satisfying the trigger is fairly flat in $\Lambda_{\text{IC}}$, but increases with increasing mass due to the slower motion of the particles.
}  
\label{fig:stopTime}
\end{figure}

Our simulation of the quirk propagation determines
$\tau_{\mbox{\scriptsize stop}}$, the time it takes for
the quirk pair to come to a stop within the detector
with a separation of $\sim1$\AA, and 
with negligible center-of-mass momentum.  
This time is 
sensitive to the confinement scale if a terminal velocity can be achieved, 
but is independent of the confinement scale for
$\Lambda_{\text{IC}}\gtrsim 300$~eV.  
This stopping time for $E'$-like quirks with $m_{E'}=500$ GeV and several choices of confinement scales is shown in Fig.~\ref{fig:stopTime}.

After the quirk pair's internal energy has been depleted, the pair will drop into a 
low-lying ``quirkonium'' bound state with a binding energy \cite{DeLuca:2018mzn},
\beq
E_{B,E'}(n) \simeq - \frac{9\alpha_{\text{IC}}(m_{E'})^2}{16n^2} m_{E'} \hspace{8mm} E_{B,T'}(n) \simeq - \frac{4\alpha_{s}^2(m_{T'})}{9n^2} m_{T'}, 
\label{eq:tau3}
\eeq
which is at most a few GeV.   
The decay of these bound states is 
prompt, with lifetimes of $\tau_{\mbox{\scriptsize decay}} \sim m_Q^{-1} \alpha^{-5}\sim 10^{-18}$~s.

    %%%%%%%%%%%%%%%%%%%%%%%%%%%%%%%%%%%%%%%%%%%%%%%%%%
 \section{Limits on Quirks\label{sec:limits}} 
 %%%%%%%%%%%%%%%%%%%%%%%%%%%%%%%%%%%%%%%%%%%%%%%%%%

Our signal comes from quirk annihilation to jets, photons, or electrons.
We assume that other modes do not contribute to our signal.
The relative probability into the major decay channels in the $T'$ model are \cite{Kang:2008ea}
 \beq
 gg : jj : {\rm IC} = \frac{32}{27} \alpha_s^2 :\frac{2 n_f}{9}\alpha_s^2 : \frac{N_{\text{IC}}^2-1}{6 N_{\text{IC}}^2} \alpha_{\text{IC}}^2,
 \eeq
  where $n_f=6$ is the number of quark species and
  \beq
  \alpha_{\text{IC}}(m) \simeq \frac{4\pi}{1+\frac{22}{3} N_{\text{IC}} \ln{\frac{4m}{\Lambda_{\text{IC}}}}}.
\eeq
 Across the parameter space of interest, this results in a subpercent level branching fraction into the invisible infracolor glueballs.
 
      \begin{figure}[t]
\begin{center}
\includegraphics[scale=0.6]{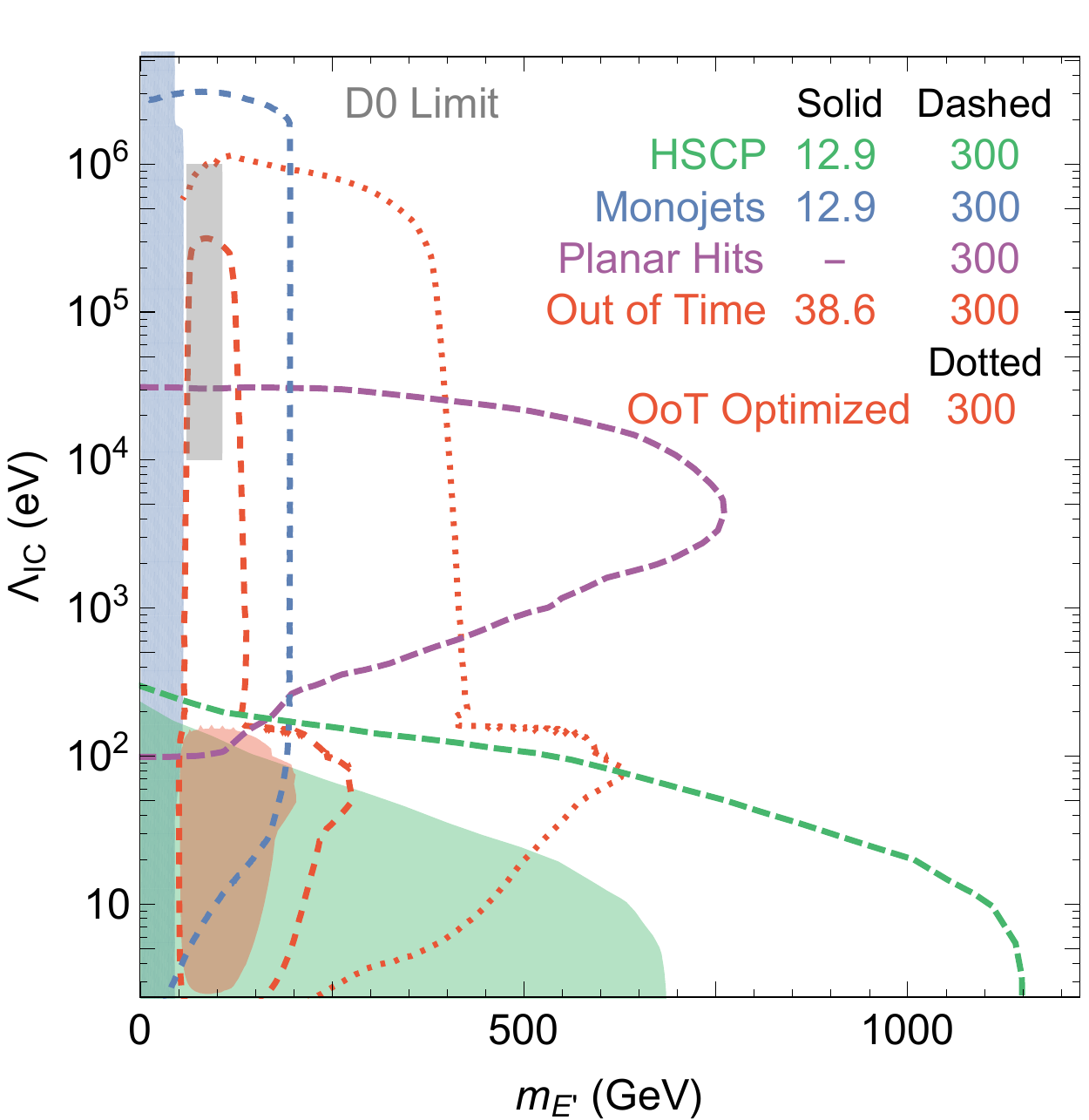} \includegraphics[scale=0.6]{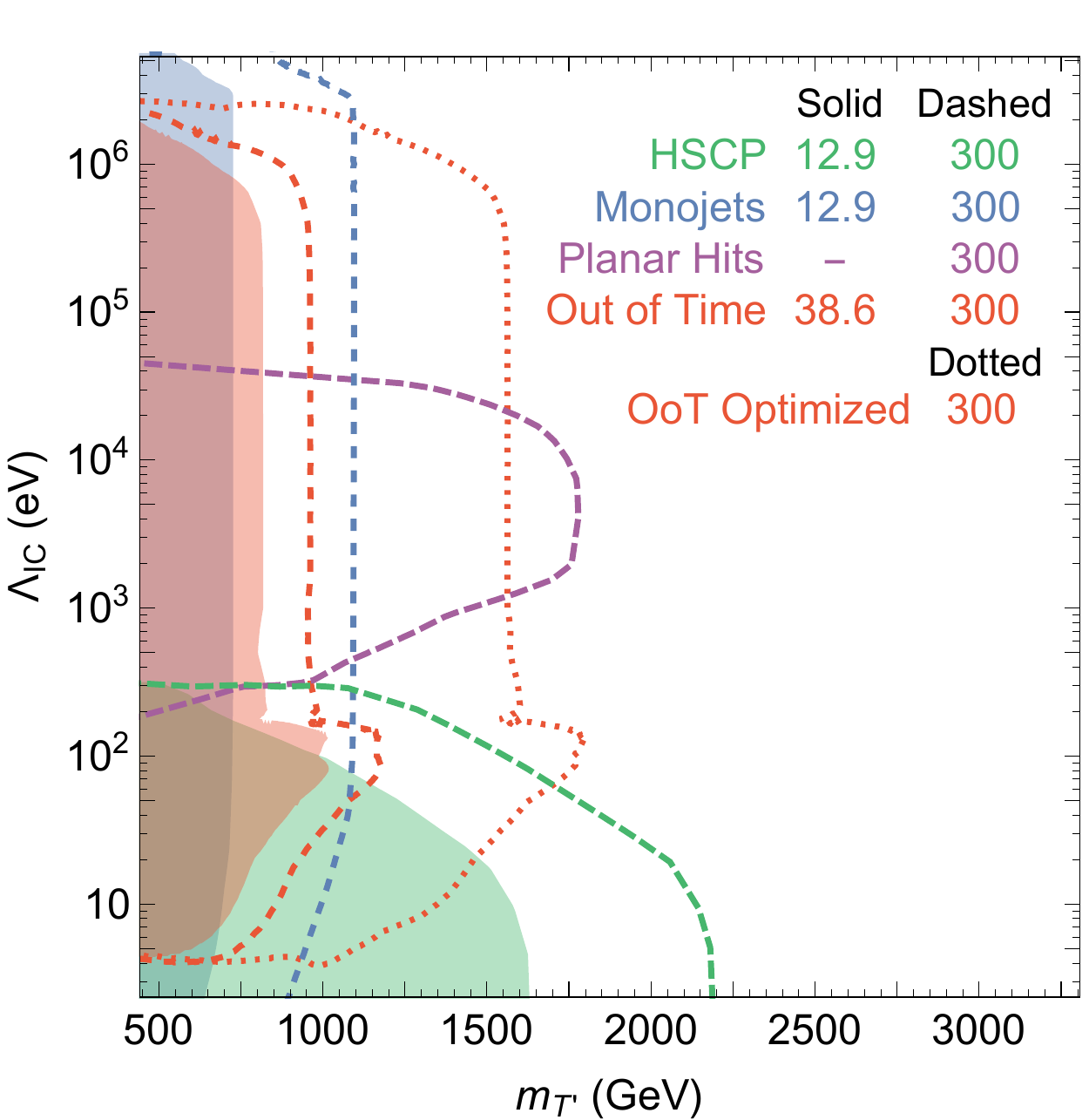}
\end{center}
\caption{{\bf Left:} Current and projected constraints on $E'$-like quirks (standard model quantum numbers of the right-handed electron $(1,1)_{1}$) with a confining SU$(2)_{\text{IC}}$ force.  {\bf Right:} Current and projected constraints on $T'$-like quirks (standard model quantum numbers of the right-handed top quark $(\bar 3,1)_{-\frac 23}$) with a confining SU$(2)_{\text{IC}}$ force.   The limits from the existing CMS out-of-time search \cite{Sirunyan:2017sbs} are shown in solid red.  Projected limits at 300 fb$^{-1}$ using a na\"ive extrapolation where an identical search strategy with the same acceptance is shown in dashed red.  A hypothetical optimization of the search where the entire calorimeter is included, $\abs{\eta}<3$, and backgrounds are reduced to zero, e.g.~via higher energy thresholds, to illustrate the possible sensitivity achievable through this method is shown with the dotted red line.   Limits from HSCP and monojet searches \cite{Farina:2017cts} are shown in green and blue, respectively.  The proposed planar hit search \cite{Knapen:2017kly} is shown in purple for 300 fb$^{-1}$.  The limit on $E'$-like quirks from D0 \cite{Abazov:2010yb} is shown in gray.}  
\label{fig:EW}
\end{figure}
 
The $E'$ model on the other hand has important ``invisible'' decays into infracolor glueballs, muons and neutrinos.  The relative probabilities are \cite{Kang:2008ea}
 \beq
 \gamma\gamma : f\bar f : \text{IC} = 2\alpha_{\text{EM}}^2 :  \alpha_{\text{EM}}^2 c_f^2 : \frac{N_{\text{IC}}^2-1}{2 N_{\text{IC}}^2}\alpha_{\text{IC}}^2
 \eeq
 where $c_f^2 \simeq (e_f + \tan\theta_W g_{Z,V}^{(f)})^2 +(\tan\theta_W g_{Z,A}^{(f)})^2$  is the combined $\gamma^*/Z^*$ coupling of the quirks to standard model fermions.  These relative branching fractions result in a 40--65\% visible decay rate across the parameter space of interest.
 
Before the quirks reach the beam pipe, they are propagating in vacuum in a state
with angular momentum $J \sim 1$.
They therefore lose energy due to infracolor glueball emission.
As shown in \cite{Kang:2008ea}, for highly excited states
the rate of glueball emission is inversely proportional to the classical crossing 
time, which we model by assuming that there is a probability $\epsilon_\text{emit}$
to emit 
 $\order{\Lambda_{\text{IC}}}$ of energy through infracolor glueballs
in one classical crossing time.
Infracolor glueball emission also changes the angular momentum of the quirk
pair by $\Delta J \sim \pm 1$, 
effectively giving rise to a random walk for the angular momentum vector in space.
Because this random walk is in 3 dimensions, it effectively 
prevents re-annihilation until the quirk pair loses all its energy.%
\footnote{Ref.~\cite{Kang:2008ea} modeled this with a 1-dimensional random
walk, and therefore concluded erroneously that there was an appreciable
probability for quirks in vacuum to annihilate before losing all their 
energy to infracolor glueball emission.}
The distance the quirk pair can travel in vacuum is therefore
\beq
L_{\text{ann}} \simeq L_\text{cross} N_\text{cross} 
\simeq \lp\beta\gamma\rp_0\lp\beta\gamma\rp_{\text{cm}} \frac{m_Q E_{\text{int}}}{\epsilon_\text{emit}\Lambda_{\text{IC}}^3},
\label{eq:Lann}
\eeq
 where  $L_\text{cross}$ is the typical lab frame length traversed in one oscillation, $N_\text{cross}$ is the number of oscillations required for the quirk pairs to lose all of their internal energy, $\gamma_0 = m_{\text{int}}/(2m_Q)$ is the boost factor of the quirks in their center-of-mass frame, $\gamma_{\text{cm}} = E_{\text{tot}}/m_{\text{int}}$ is the boost factor of the quirk system in the lab frame (see eq \ref{eq:Edist}).
 For simplicity, we assume that $\epsilon_{\text{emit}} = 0.1$ and require that the quirk pair reaches the beam pipe (1 cm transverse) prior to saturating the annihilation length.\footnote{It is possible that some of our quirks which in simulation did not stop in the fiducial region of the detector now would due to the additional energy loss prior to reaching the beam pipe. This effect should be qualitatively unimportant and subdominant to the sensitivity of changing the unknown $\epsilon_{\text{emit}}$ parameter.}
 
The CMS search requires a calorimeter jet with $E>70$ GeV and $\abs{\eta}<1.0$ at least 2 bunch crossings away from pp collisions.   To better reject cosmic muon backgrounds, CMS imposes a veto on a variety of drift tube (DT) patterns (12.3\% of events).  To reject beam halo muon backgrounds, the CMS search additionally vetoes events with any cathode strip chamber (CSC) segments having at least five reconstructed hits (5.6\% of events).   We include both of these vetoes as a fixed percentage reduction of our signal efficiency.

While the reconstruction efficiency for visibly decaying quirks, which give two back-to-back high-energy objects, is expected to be a bit better than for $R$-hadrons, where half of the energy goes invisibly, we conservatively use the energy dependent function for the gluino from Ref.~\cite{Sirunyan:2017sbs} that quickly asymptotes to 55\% for our masses of interest.

 In Fig.~\ref{fig:EW}, we show in shaded red the limits from the out-of-time search for $E'$-like and $T'$-like quirks (left and right, respectively).  The recast limits from monojet and HSCP searches \cite{Farina:2017cts} are shown in blue and green, respectively.\footnote{For the monojet limits, we estimate the reduction in sensitivity at high confinement scales due to annihilations within the beampipe Eq.~(\ref{eq:Lann}). The monojet constraints die off at larger confinement scales than the stopped particle limits as the fraction of the sample that meets the monojet requirements are typically more boosted and more likely to escape the beampipe.  We assume that limits on quirks below $m_Z/2$ remain robust.  Any modifications to the efficiency due to the presence of a straight, highly-ionizing track have not been addressed.} For $E'$-like quirks, we include the limit from the D0 (in gray) \cite{Abazov:2010yb}.   Via dashed lines of the same colors we show projections to 300 fb$^{-1}$, adding (in purple) the projections for the planar hit reconstruction proposal \cite{Knapen:2017kly}.   We note that the dashed red lines illustrating the out-of-time decay search assume both the search and bunch train patterns remain identical in the larger dataset. The dotted red lines assume zero background and expands the calorimeter range out to $\abs{\eta}\leq 3$, which should be viewed as a realistic achievable sensitivity under some simple modifications to improve the stopped particle search that will be discussed in the next section.
 
 %%%%%%%%%%%%%%%%%%%%%%%%%%%%%%%%%%%%%%%%%%%%%%%%%%
 \section{Improving Sensitivity to Quirks\label{sec:improve}}
 %%%%%%%%%%%%%%%%%%%%%%%%%%%%%%%%%%%%%%%%%%%%%%%%%%
 
At the LHC, the kinematics of quirks are very different than those of $R$-hadrons.
In this section, we will discuss several simple changes to the search strategy that 
can greatly improve the sensitivity to quirks.

\begin{figure}[t]
\begin{center}
\includegraphics[scale=1]{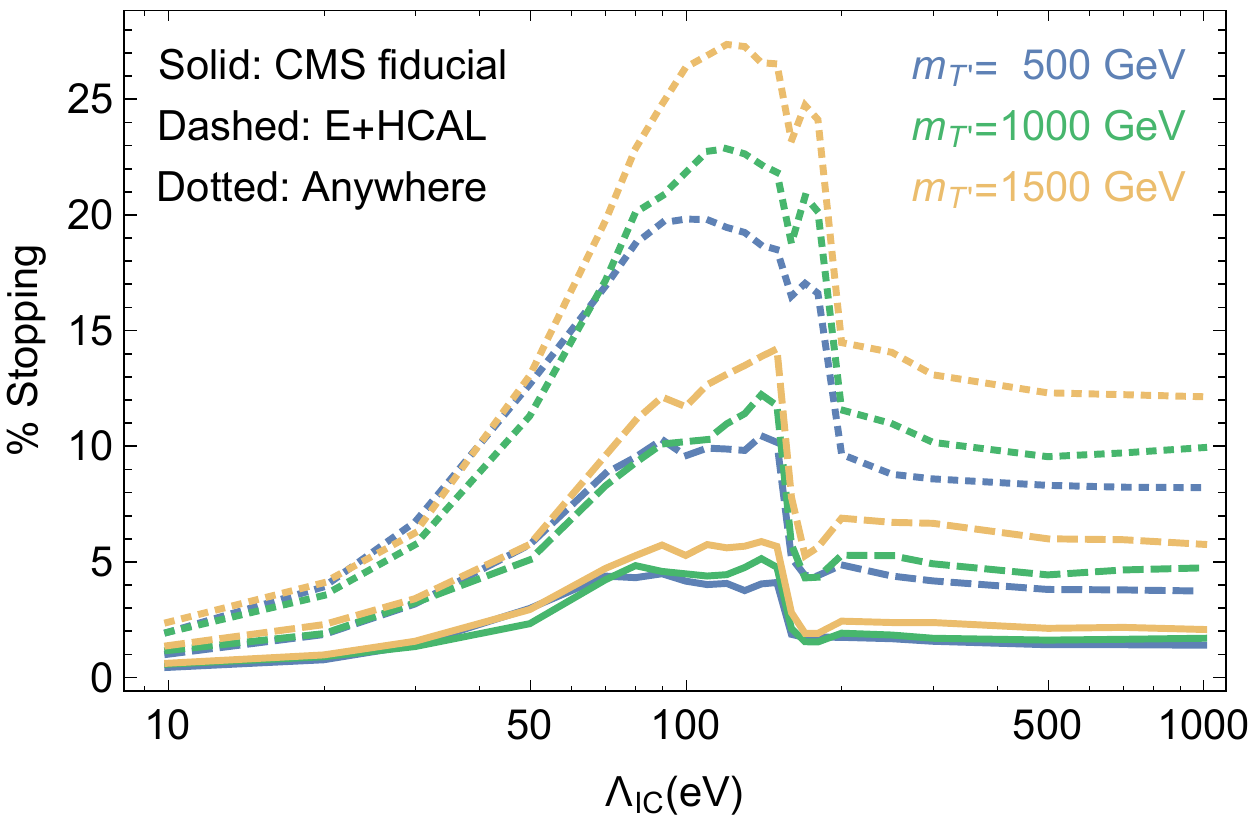}
\end{center}
\caption{Percentage of $T'$-like quirks that stop in CMS for a given confinement scale.  The solid lines show the fraction of quirks that stop in the CMS fiducial region ($\abs{\eta}<1.0$).   Dashed lines display the fraction that stops in the entire calorimeter out to $\abs{\eta}<3.0$.  Dotted lines display the fraction of quirks that stop anywhere in the CMS detector (mostly in the muon system).  As the mass of the quirks increases, the fraction that stops also increases.}  
\label{fig:stoppingFraction}
\end{figure}

First, the quirk pair system tends to be produced at very high $\eta$.  
This is because any nonzero $p_T$ of the quirk pair requires that it recoil against
another hard object, typically a jet.
Generally, the distribution peaks around $\eta \sim 3$, achieving higher values for 
lower quirk masses, while colored production tends to be a little less forward.  
This is in contrast to $R$-hadrons, which are produced centrally.
The $\abs{\eta}<1.0$ cut applied in the CMS search is heavily impacting the signal efficiency.  While it is also true that quirks in the far forward region will often have a higher $E_{\text{cm}}$  (reducing the likelihood that they will stop), the net effect of including the full calorimeter is about a factor of three enhancement in signal acceptance, see Fig.~\ref{fig:stoppingFraction}.  Additionally, quirks that stop in the forward region tend to take longer to stop due to the additional distance traversed at rather low velocities, which make them more likely to fall within a triggerable window at large confinement scales.   On the other hand, beam halo backgrounds are expected to be much larger in this region, and the CSC veto used at CMS may no longer be sufficient.   Imposing a large enough energy threshold for the signal in this region may be sufficient to suppress these backgrounds, but detailed experimental study would be required.

 Second, while the visible energy emitted in an $R$-hadron decay is expected to be $E\lesssim \frac{m_R}{2}$, annihilating quirks deposit $E\sim 2m_Q$ into the detector.  Especially in the case of $T'$-like quirks, this exceptionally high energy deposit should improve available triggering options.  In other words, $\sim 2$ TeV of energy appearing in a few cells of the calorimeter is typically sufficient for triggering purposes even within an active bunch window.  Additionally, as evidenced by the ATLAS stopped particle search at 8 TeV \cite{Aad:2013gva}, the use of signal regions with higher energy thresholds greatly reduces beam halo backgrounds and substantially reduces cosmic muon backgrounds to the point that no events were observed in the signal region with a 300 GeV cut applied to the leading jet energy.  Expanding the existing analysis to include multiple signal regions with increasing energy thresholds for background rejection and wider $\eta$ ranges for enhanced acceptance would potentially allow for a nearly background free search with a very large signal acceptance for quirks.  In Fig.~\ref{fig:EW}, we show via a dotted red line the sensitivity that would be obtained in 300 fb$^{-1}$ of data with the calorimeter range expanded to $\abs{\eta}<3$ and assuming no background.  Any possible additional timing modifications that would affect the fraction of quirks annihilating within a triggerable window have not been included.   While the background-free assumption may be overly aggressive, especially in the case of $E'$-like quirks, this curve is meant to illustrate what could potentially be achieved via reasonable modifications to the existing search strategy.
 
  Additionally, CMS is constructing a new endcap calorimeter ($1.5<\abs\eta<3.0$) that would have $\sim$25 ps timing resolution and individually instrumented layers allowing for finely-grained 4d shower reconstruction \cite{Collaboration:2293646}.   The capabilities of this instrument are perfect for detecting out-of-time annihilation of quirks.   It also has the potential to use the unique evolution pattern of annihilating quirks for both triggering at L1 and for reliably distinguishing the quirks from all backgrounds.  Moreover, as collisions typically happen only within the first 2.5~ns of a bunch crossing (and the overwhelming bulk of these collisions within a much smaller window of a few hundred ps), this instrument could potentially be used to capture exceptionally large, forward calorimeter energy deposits that are not offset by $d/c$ in time from any collisions, potentially allowing the percent of quirks that would decay within a triggerable window to be $\order{90\%}$.

 Lastly, while many quirks stop within the calorimeters,  approximately the same number again are expected to stop in the dense iron return yoke of the muon system,  see Fig.~\ref{fig:stoppingFraction}.  Especially in the case of heavy $T'$-like quirks, back-to-back TeV jets appearing in the drift tubes would look startlingly different than most backgrounds.   While we make no effort to construct a search strategy around this possibility, we note that it would be an interesting option for CMS to consider in order to maximize sensitivity to heavy colored quirks.  Unlike at CMS, the prospects for detecting quirks that stop within the more sparsely instrumented ATLAS muon chamber are slim.  
   
   %%%%%%%%%%%%%%%%%%%%%%%%%%%%%%%%%%%%%%%%%%%%%%%%%%
 \section{Discussion\label{sec:discussion}} 
 %%%%%%%%%%%%%%%%%%%%%%%%%%%%%%%%%%%%%%%%%%%%%%%%%%
 
   In this work we recast the CMS search for the out-of-time decay of stopped particles and apply it to the out-of-time annihilation of stopped quirks.  We find that possible modifications to the search strategy could make an out-of-time search the most effective way of accessing quirks across a very wide range of confinement scales.   While multiple assumptions were made throughout this work to facilitate our simulation and recasting, these assumptions should not have a qualitative impact on the basic search improvements proposed here.   What is presented here should be superseded by a proper treatment within a sophisticated simulation framework (e.g., GEANT4 \cite{Agostinelli:2002hh}) that is approved by an experimental collaboration.
      
   While the existing stopped particle search places constraints competitive with those from monojet searches, a few straightforward modifications to the search strategy used in out-of-time searches could greatly enhance the sensitivity to quirks.  The unique kinematics that manifest in quirky systems strongly encourage searches to probe higher pseudorapidities and use higher energy thresholds to beat down backgrounds.  Several of the modifications proposed here could allow for out-of-time searches to become the discovery mode for quirks over a large range of confinement scales, whereas in the case of $R$-hadrons this approach is currently subdominant to the sensitivity obtained with heavy stable charge particle searches.   Additionally, in the event an excess were seen in monojets,  identifying this excess as originating from quirks rather than the plethora of other possibilities, e.g., dark matter, would rely on either a dedicated search or a complementary excess in a different channel (such as an out-of-time search akin to the one recast in this work).  
   
 Although this study focused entirely on the 13 TeV CMS study, ATLAS, which to date only has 7 and 8 TeV results, would be expected to be able to achieve similar sensitivity.  The cuts used in the existing searches at the two experiments have some important differences.  ATLAS uses a slightly wider central range of $\abs\eta<1.2$ with a slightly lower jet energy threshold of 50 GeV.   However, they require the signal to appear at least six bunch crossings after all filled selections, which for quirks would result in a substantially lower probability to decay within an acceptable trigger window.  To reduce background from detector noise and muons, events where fewer than four calorimeter cells contain 90\% of  the leading jet energy are vetoed, the leading jet must have a $p_T$ weighted width $\Delta R>0.04$, and at least 50\% of the leading jet's energy must be recorded in the barrel hadronic calorimeter of ATLAS.   Together, these other important cuts reduce the reconstruction efficiency of their gluino samples considerably relative to CMS.  On the other hand, ATLAS has multiple bins with increasing energy thresholds, allowing them to reduce their background, even down to zero observed events in their highest considered energy bin.
   
  One assumption made in projecting the future reach for these out-of-time searches is that the fraction of quirks decaying within a triggerable window will not change substantially from the 2016 fill schemes.  This assumption is particularly bold as LHC running has switched dominantly to an 8--4 scheme where eight colliding bunches are followed by four empty bunches.   The incarnation of the CMS search recast for this study, which requires the signal to be at least two bunch crosses away from an active bunch, would have an extremely reduced sensitivity to this fill scheme, forcing it to rely entirely on the abort gap for sensitivity.  It is important that the search strategy evolves to accommodate this new scheme without drastically reducing the trigger live time.   New technologies with precision timing may be able to assist in expanding this live time further.
   
 The out-of-time searches at ATLAS and CMS have the potential to probe quirks across a very large confinement scale range.  This approach is complementary to the HSCP searches and could be much more sensitive than monojet searches if the simple modifications to the approach discussed in this work were implemented.   Moving forward, we encourage both ATLAS and CMS to consider quirks as a benchmark in their out-of-time decay searches, and to design their signal regions to target these truly exotic particles.

\begin{acknowledgments}
  We thank Juliette Alimena, Gordon Baym, Yuri Gershtein, Simone Pagan Griso, Roni Harnik, Simon Knapen, Ted Kolberg, Zhen Liu, and Yuhsin Tsai for useful discussions. We thank Christina Gao for collaboration in the early stages of this work.  We are grateful to Simon Knapen for valuable comments on the manuscript.  JAE is partially supported by the NSF CAREER grant PHY-1654502 and DOE grant DE-SC0011784.  MAL was supported by DOE grant DE-FG02-91ER406746.  JAE thanks the Aspen Center for Physics, which is supported by National Science Foundation grant PHY-1607611, where part of this work was completed.
  \end{acknowledgments}
  
  \appendix
  
  %%%%%%%%%%%%%%%%%%%%%%%%%%%%%%%%%%%%%%%%%%%%%%%%%%
   \section{The CMS Detector\label{sec:CMS}} 
   %%%%%%%%%%%%%%%%%%%%%%%%%%%%%%%%%%%%%%%%%%%%%%%%%%
   
      \begin{table}
\begin{center}
\begin{tabular}{|c|ccc|cc|cc|}
\hline
Component & Material &$\rho_m$ & $\zeta$ & $R_{\text{min}}$ & $R_{\text{max}}$ & $\abs{Z_{\text{min}}}$ & $\abs{Z_{\text{max}}}$ \\
\hline\hline
Tracker & Silicon & 2.33 & 0.05 & 2.2 &118.5 & 0 & 293.5 \\
\hline
ECal  & Lead Tungstate & 8.30 & 0.33 & 118.5 & 181.1 & 0 &390.0 \\
  &  (PbWO$_4$) &&   & $\eta =3$ & 181.1 & 293.5&390.0 \\
 \hline
HCal  & Copper &  8.96 & 0.68 & 181.1 & 286.4 & 0 &568.0 \\
  & &&   & $\eta =3$ & 286.4 &390.0 & 568.0 \\
 \hline
Magnet & Iron & 7.87 & 0.5 & 295.0 & 380.0 & 0 &645.0 \\
 \hline
MS & Iron &  7.87 & 1 & 380 & 410 & 0 & 140.0 \\
 barrrel & & &   & 455 & 490.5 & 0 & 724.0 \\
  & &&   & 528.5 & 597.5 & 0 & 661.0 \\
  & & & & 635.5 & 700 & 0 & 661.0 \\
  \hline
 MS  & Iron & 7.87 & 1   & $\eta =3$ & 105 & 568 & 630 \\
 endcap & & &   & $\eta =3$ & 270 & 630 & 724 \\
  & &  & & $\eta =3$ & 695.5 & 724 & 791.5 \\
  & & &  & $\eta =3$ & 129 & 791.5 & 849.5 \\
  & & &  & $\eta =3$ & 695.5 & 849.5 & 917 \\
  & & &  & $\eta =3$ &140 & 917 & 975 \\
  & & &  & $\eta =3$ & 695.5 & 975 & 1005 \\
  & & &   & $\eta =3$ &150 & 1005 &1063 \\
  & & &  & $\eta =3$ &695.5 & 1063 & 1083 \\
  \hline
\end{tabular}
\end{center}
\caption{Details of the cylindrical CMS detector used in this study. $\rho_m$ is in g/cm$^3$, all $R$ and $Z$ dimensions are in cm.  An $R_{\text{min}}$  value of $\eta =3$ implies that for a given value of $Z$, the inner radius is set where $\eta=3$.  Outside of the beampipe vacuum, everywhere not mentioned in this table, notably the muon system drift chambers, is assumed to consist of air, $\rho_{\text{air}} = 0.001165$ g/cm$^3$ and $\zeta = 1$. \label{tab:CMS}}
\end{table}

      \begin{figure}[t]
\begin{center}
\includegraphics[scale=1]{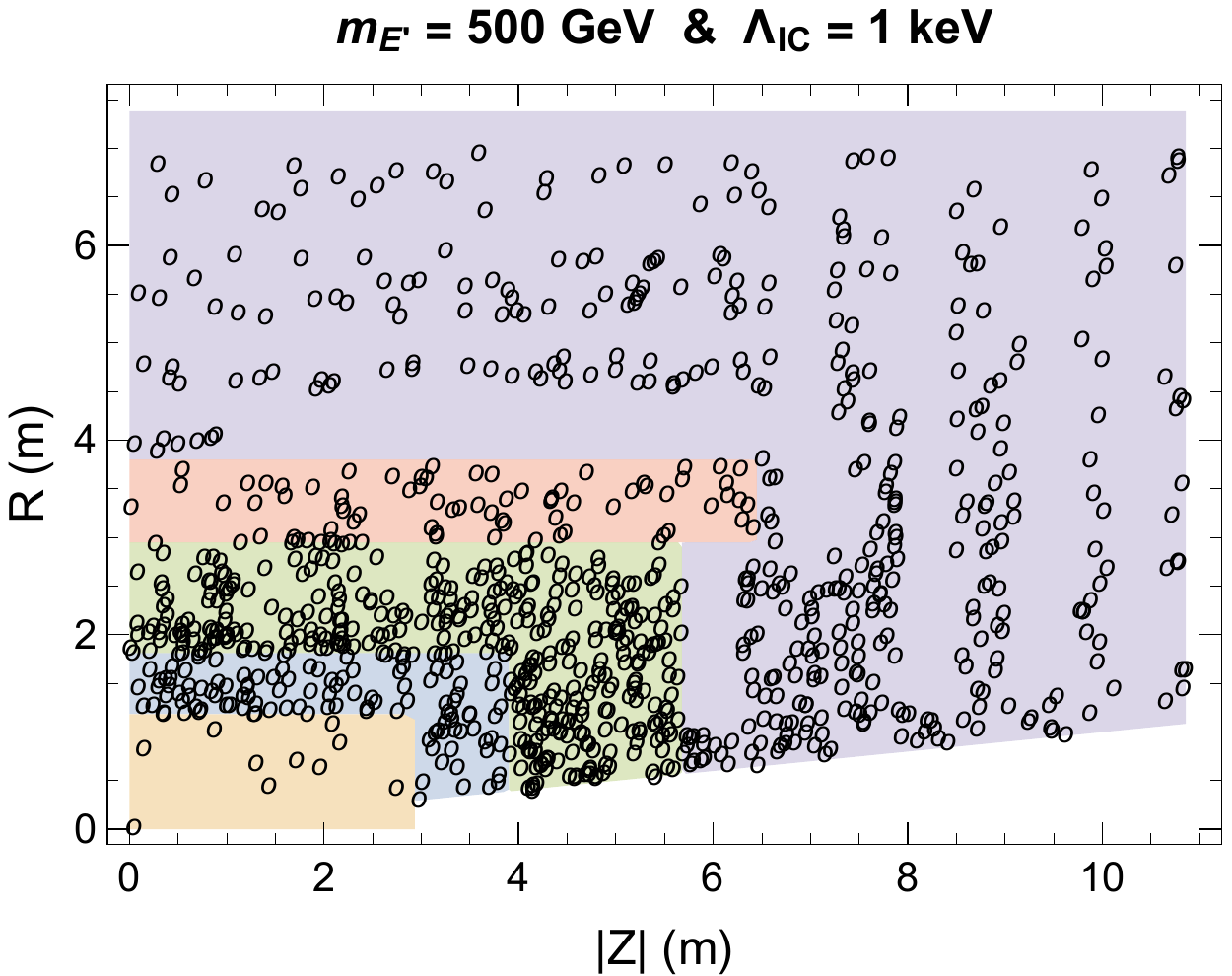}
\end{center}
\caption{Positions where 983 $E'$-like quirks with $m_{E'}=500$ GeV and $\Lambda_{\text{IC}}=1$~keV stop in the CMS detector within a quadrant of $(R,\abs Z)$.   Moving outwards we have the tracker (gold), ECal (blue), HCal (green), magnet (red), and muon system (purple).  The most dense systems (HCal, MS iron return yokes) stop the most particles, while the least dense (tracker, MS drift tubes) stop very few.   Many quirks are seen to stop at high values of $\eta$,  motivating an extension of the calorimeter $\eta$ range used in the out-of-time searches.}  
\label{fig:StoppingPos}
\end{figure}

We approximate the CMS detector as being cylindrically symmetric.   As we are not concerned about the specific location of individual hits, but rather the overall energy loss, we break the detector into distinct zones of a uniform medium of the material in question.   
The Bethe-Bloch function can then be written
\beq
\left\langle \!\frac{dE}{dx}\!\right\rangle\!(\vec x_i,v_i) \simeq \zeta(\vec x)\rho_m(\vec x)\left\langle\!\frac{dE}{dx}\!\right\rangle\!\lp\beta \gamma\rp,
\eeq
where most of the dependence on the material is contained in the density $\zeta\rho_m$, where $\rho_m$ is the density of a solid of the material in question (either silicon, lead tungstate, copper [equivalent to brass], or iron) and $\zeta(\vec x)\leq1$ is a numerical factor extracted from material budget studies where available \cite{CMS:1997ysd,Bayatian:922757,Amsler:2009ova}, and approximated elsewhere.  There is some slight dependence of $\vev{dE/dx}$ on the material \cite{Patrignani:2016xqp} that is included in our simulation, but the majority of the material specific dependence is contained in the density.  The layout of CMS is approximated as a series of coaxial cylindrical shells shown in Table~\ref{tab:CMS} \cite{Chatrchyan:2009hg}.  In Fig.~\ref{fig:StoppingPos}, we show the positions where simulated $E'$-like quirks with $m_{E'}=500$ GeV and $\Lambda_{\text{IC}}=1$~keV stop in the CMS detector.  At this confinement scale, there are no regions within the detector where the quirks achieve a terminal velocity.

  %%%%%%%%%%%%%%%%%%%%%%%%%%%%%%%%%%%%%%%%%%%%%%%%%%
   \section{Energy Loss at the Atomic Scale \label{sec:Eloss}} 
   %%%%%%%%%%%%%%%%%%%%%%%%%%%%%%%%%%%%%%%%%%%%%%%%%%

Once the size of the quirk bound state falls below
$\text{1~\AA}~\sim 1/{\text{keV}}$,
the quirks can no longer be viewed as independently ionizing objects. 
The quirks dominantly stop in the brass material in the hadronic calorimeter, and we
will assume that when the quirk string length is of order 1~\AA\ the quirk
pair is bound to the material and no longer moves.
In order to annihilate, the quirk pair must loose the remainder of its
energy and angular momentum.
In this appendix, we consider several energy loss mechanisms that 
may be important in this regime:
radiation of infracolor glueballs,
electromagnetic radiation,
and induced currents in the material due to the oscillating electric
dipole of the quirks.

%%%%%%%%%%%%%%%%%
\subsection{Infracolor Glueball Emission}
The energy lost through the non-perturbative emission of infracolor glueballs is important when the impact parameter $b\lesssim \Lambda_{\text{IC}}^{-1}$.  Impact parameters of this size are achieved when angular momentum is not too high 
\beq
J \lesssim 10^4 \lp\frac{m_Q}{\mbox{TeV}}\rp^{1/2}.
\eeq
This can be compared to the maximum angular momentum at 1 \AA\ of
\beq
J_{\text{max}}(r = \text{\AA}) \sim 10^4 \lp\frac{m_Q}{\mbox{TeV}}\rp^{1/2} \lp\frac{\Lambda_{\text{IC}}}{\mbox{keV}}\rp.
\eeq
Just as the forward motion of a quirk pair is drained more quickly than the internal motion, once the system slows, acquired angular momentum is drained much more quickly than the internal energy, so that $J\ll J_{\text{max}}$ 
appears to be
generic by the time the quirks are within one \AA\ of each other.  On the other hand, this is in part an artifact of 
approximating the ionization energy loss as a classical friction.  
A full simulation must take into account the fact that the individual ionization
interactions will give the quirks a momentum transfer transverse to their
motion that cancels out only on average.  
We have not attempted to model this effect.
We can parameterize the emission of infracolor glueballs by assuming that at each classical crossing, the quirks have a probability $\epsilon_{\text{emit}}$ to emit $\Lambda_{\text{IC}}$ of energy into infracolor glueballs, so that 
\beq
\frac{dE}{dt} = - \frac{\epsilon_{\text{emit}}\Lambda_{\text{IC}}}{T}.
\label{eq:dEdt}
\eeq
If $\Lambda_{\text{IC}}\gtrsim\,$\AA$^{-1}$, the potential begins in a linear regime $V(r)\sim \Lambda_{\text{IC}}^2 r$, where the quirks experience a constant acceleration, $a = \Lambda_{\text{IC}}^2 m_Q^{-1}$.  The classical crossing time is $T\simeq \frac{2v}{a} =2 \Lambda_{\text{IC}}^{-2}\sqrt{2m_Q E}$, where we substituted $v= \sqrt{2 E/m}$.    Integrating Eq.~(\ref{eq:dEdt}), we find the time required to lose $\Delta E \sim \Lambda_{\text{IC}}^2$\AA\ of energy stored in the string in order to enter the Coulombic region of the potential is
\beq
\tau_{\mbox{\scriptsize linear}} \sim \frac{4\sqrt{2 m_Q} \lp\Delta E\rp^{3/2}}{3 \epsilon_{\text{emit}} \Lambda_{\text{IC}}^3} \simeq 10^{-13}~\text{s}\lp\frac{m_Q}{\mbox{\small TeV}}\rp^{1/2} \lp\frac{0.1}{\epsilon_{\text{emit}}}\rp  
\hspace{10mm} \Lambda_{\text{IC}}\gtrsim\text{\AA}^{-1},
\label{eq:tau1}
\eeq 
which is independent of $\Lambda_{\text{IC}}$.  Once the quirk pair has shrunk to $r\lesssim\Lambda_{\text{IC}}^{-1}$, the bound state is in the Coulombic regime, where $V(r)\sim -\alpha_{\text{IC}}(r)/r$ and the classical crossing time (half a period) is dictated by Kepler's law, $T=\pi \sqrt{m_Qr^3\alpha_{\text{IC}}^{-1}}$. From Eq.~(\ref{eq:dEdt}), we can express the change in the binding energy $B =-V(r)$ as
\beq
\frac{dB}{dt} = \frac{\epsilon_{\text{emit}}\Lambda_{\text{IC}}B^{3/2}}{\pi \alpha_{\text{IC}} \sqrt{m_Q}}.
\label{eq:tau2diffeq}
\eeq  
 Again integrating and noting that the initial binding energy $B_i \sim \Lambda_{\text{IC}}$ is much smaller than the final binding energy, $B_0\sim \alpha_{\text{IC}}^2 m_Q$, the time to drop to the ground state is approximately
\beq
\tau_{\mbox{\scriptsize Coulomb}}  \simeq \frac{2\pi\alpha_{\text{IC}} \sqrt{m_Q}}{\epsilon_{\text{emit}}\Lambda_{\text{IC}}^{3/2}}\simeq 10^{-12}~\text{s}\lp\frac{m_Q}{\mbox{\small TeV}}\rp^{1/2} \lp\frac{\mbox{\small~keV}}{\Lambda_{\text{IC}}}\rp^{3/2}\lp\frac{0.1}{\epsilon_{\text{emit}}}\rp \hspace{10mm} \Lambda_{\text{IC}}\gtrsim\text{\AA}^{-1},
\label{eq:tau2}
\eeq
where we have used that $\alpha_{\text{IC}}(\Lambda_{\text{IC}})\sim 1$.    On the other hand if $\Lambda_{\text{IC}}\lesssim\,$\AA$^{-1}$, the quirks have already dropped into the Coulombic regime when $r\lesssim 1$~\AA\ (and $\tau_{\mbox{\scriptsize linear}}=0$).  The same calculation as before Eq.~(\ref{eq:tau2diffeq}) can be used, but now with the initial binding energy moved from $\Lambda_{\text{IC}}\to \alpha_{\text{IC}}(\text{\AA}^{-1})\text{\AA}^{-1}$,
 \beq
\tau_{\mbox{\scriptsize Coulomb}} \simeq \frac{2\pi \sqrt{\alpha_{\text{IC}}m_Q\text{\AA}}}{\epsilon_{\text{emit}}\Lambda_{\text{IC}}}\simeq 10^{-12}~\text{s}\lp\frac{m_Q}{\mbox{\small TeV}}\rp^{1/2} \lp\frac{\mbox{\small~keV}}{\Lambda_{\text{IC}}}\rp\lp\frac{0.1}{\epsilon_{\text{emit}}}\rp \hspace{10mm} \Lambda_{\text{IC}}\lesssim\text{\AA}^{-1}.
\eeq
 Across our parameter range of interest, $\tau_{\mbox{\scriptsize linear}}+\tau_{\mbox{\scriptsize Coulomb}}\lesssim 1$~ns unless $\epsilon_{\text{emit}}$ is very small.   However, this depends crucially on the impact parameter being small enough for infracolor glueball emissions to happen at all.   While for $\Lambda_{\text{IC}}\lesssim $~keV, this is almost certain to be the initial state once the quirks stop, for $\Lambda_{\text{IC}}\sim $ MeV, it seems less likely.  Once the size of the quirk pair drops into the Coulomb regime $b < r \sim \Lambda_{\text{IC}}^{-1}$, the emission of infracolor glueballs should become an effective way to radiate the remaining energy. 

%%%%%%%%%%%%%%%%%
\subsection{Electromagnetic Radiation}
The quirks form an oscillating dipole that radiates away electromagnetic energy.
In estimating this effect, it is important to take into account the effects
of the conductor on electromagnetic radiation.
We will use the Drude model, which assumes that conduction electrons can be
described by drifting electrons with a phenomenological damping force due to
collisions of the electrons with the lattice.
That is, the current density in the conductor is given by
\beq
\vec{J} = -e n_e \vec{v}_d,
\eeq
where the drift velocity $\vec{v}_d$ satisfies the equation of motion
\beq
\frac{\partial \vec{v}_d}{\partial t} 
= - \frac{1}{\tau_c} \vec{v}_d
- \frac{e}{m_e} 
(\vec{E} + \vec{v}_d \times \vec{B}).
\eeq
The phenomenological collision time is $\tau_c \sim 10^{-14}$~s in copper,
and we assume this value holds for the brass in the EM calorimeter.
The effects of the magnetic field will be negligible for fields produced by
non-relativistically moving quirks, and so we have
\beq
\frac{\partial \vec{J}}{\partial t} = -\frac{1}{\tau_c} (\vec{J} - \sigma \vec{E}),
\label{eq:Drude}
\eeq
where
\beq
\sigma = \frac{e^2 n_e \tau_c}{m_e}
\eeq
is the conductivity of the conductor.
Electromagnetic plane waves in this material have the complex dispersion relation
\beq
k^2 = \frac{\omega^2}{c^2} + \frac{i \mu \sigma}{1 - i \omega \tau_c} \omega.
\eeq
For $\omega \ll 1/\tau_c$, this is approximately%
\footnote{Note that the phase velocity is larger than $c$
\[
v_p^2 = \frac{\omega^2}{k^2} = c^2 \left( 1 + \frac{\mu \sigma}{\tau_c k^2} \right),
\]
but the group velocity is
\[
v_g = \left( \frac{\partial k}{\partial \omega} \right)^{-1}
= c^2 \frac{k}{\omega}
= \frac{c}{1 + \mu \sigma / (\tau_c k^2)} < c.
\]}

\beq
k^2 = \frac{\omega^2}{c^2} + i \mu \sigma \omega
= \frac{\omega(\omega + i / \bar{\tau})}{c^2},
\eeq
where 
\beq
\bar\tau = \frac{\epsilon}{\sigma}.
\eeq
For typical metals, $\bar\tau \sim 10^{-18}~\text{s} \ll \tau_c$,
and since $1/ \bar\tau \gg 1/\tau_c \gg \omega$,
$k^2$ is nearly pure imaginary.
This means that the real and imaginary parts of $k$ are nearly equal,
and the electromagnetic wave is attenuated on a length scale equal
to its wavelength.
In the opposite limit, $\omega \gg 1/\tau_c$ we have
\beq
k^2 = \frac{1}{c^2} \left( \omega^2 - \frac{1}{\tau_c \bar\tau} \right).
\eeq
For $\omega\gtrsim 10^{16}$ s$^{-1}$, $k$ is nostly real,
so that EM waves will propagate freely in the material.
This corresponds to confinement scales of $\Lambda_{\text{IC}}\gtrsim 100$ keV.  
Also, for photons with $E_\gamma\gtrsim10$ eV ($\Lambda_{\text{IC}}\gtrsim 30$ keV), 
the photons are high enough energy to eject bound electrons in brass via the 
photoelectric effect.  
We conservatively include EM energy loss only for confinement scales above 100 keV.

The frequency of quirk oscillation, and thus of the emitted photons, is 
\beq	
\omega \sim \frac{\pi}{T} \sim \frac{\Lambda_\text{IC}}{\sqrt{2 R m_Q}}
\sim10^{14}~\text{s}^{-1}
\left( \frac{\Lambda_\text{IC}}{\text{keV}} \right)
\left( \frac{m_Q}{\text{TeV}} \right)^{-1/2},
\eeq
so for confinement scales $\Lambda_{\text{IC}}\gtrsim$ keV $\approx$ \AA$^{-1}$, 
the energy lost via radiation of photons can be described using
the Larmor formula \cite{Kang:2008ea,Burdman:2008ek},
\beq
\frac{dE_\gamma}{dt} = \frac{16\pi}{3} \alpha_{\text{EM}}  a^2.
\eeq
Since $\Lambda_{\text{IC}}\gtrsim\,$\AA$^{-1}$, the potential begins in a linear regime $V(r)\sim \Lambda_{\text{IC}}^2 r$, where the quirks experience a constant acceleration, $a = \Lambda_{\text{IC}}^2 m_Q^{-1}$.  The time required to lose $\Delta E \sim \Lambda_{\text{IC}}^2$\AA\ of energy stored in the string in order to enter the Coulombic region of the potential is
\beq
\tau_{\mbox{\scriptsize linear}} \sim \frac{3 m_Q^2\text{\AA}}{16\pi\alpha_{\text{EM}} \Lambda_{\text{IC}}^2} \sim 2~\text{s} \lp\frac{m_Q}{1\mbox{ \small TeV}}\rp^2 \lp\frac{1\mbox{ \small~keV}}{\Lambda_{\text{IC}}}\rp^2  
\label{eq:tau1}
\eeq 
Once the quirk pair has shrunk to $r\lesssim\Lambda_{\text{IC}}^{-1}$, the bound state is in the Coulombic regime, where $V(r)\sim -\alpha_{\text{IC}}(r)/r$. The acceleration can be written in terms of the binding energy $B =-V(r)$, giving
\beq
\frac{dE}{dt} = \frac{dB}{dt} \sim \frac{16\pi\alpha_{\text{EM}} B^4}{3 \alpha_{\text{IC}}^2 m_Q^2}.
\eeq  
The time to drop to the ground state ($B_0\sim \alpha_{\text{IC}}^2 m_Q$) is
\beq
\tau_{\mbox{\scriptsize Coulomb}} \sim \int^{ \alpha_{\text{IC}}^2 m_Q}_{\Lambda_{\text{IC}}} 
\frac{dB}{dB/dt}
\sim \frac{3 m_Q^2}{16\pi\alpha_{\text{EM}}}\int^{ \alpha_{\text{IC}}^2 m_Q}_{\Lambda_{\text{IC}}} \frac{\alpha_{\text{IC}}^2}{B^4}.
\label{eq:tau2}
\eeq
As this integral is dominated by small $B$, we can use that $\alpha_{\text{IC}}(\Lambda_{\text{IC}})\sim 1$ to approximate 
\beq
\tau_{\mbox{\scriptsize Coulomb}} \sim \frac{m_Q^2}{16\pi\alpha_{\text{EM}}\Lambda_{\text{IC}}^3} \simeq 2~\text{s} \lp\frac{m_Q}{1\mbox{ \small TeV}}\rp^2 \lp\frac{1\mbox{ \small~keV}}{\Lambda_{\text{IC}}}\rp^3 ,
\eeq
 which is always less than $\tau_{\mbox{\scriptsize linear}}$.   

%%%%%%%%%%%%%%%%%
\subsection{Induced Currents}

\begin{figure}[t]
\begin{center}
\includegraphics[scale=0.5]{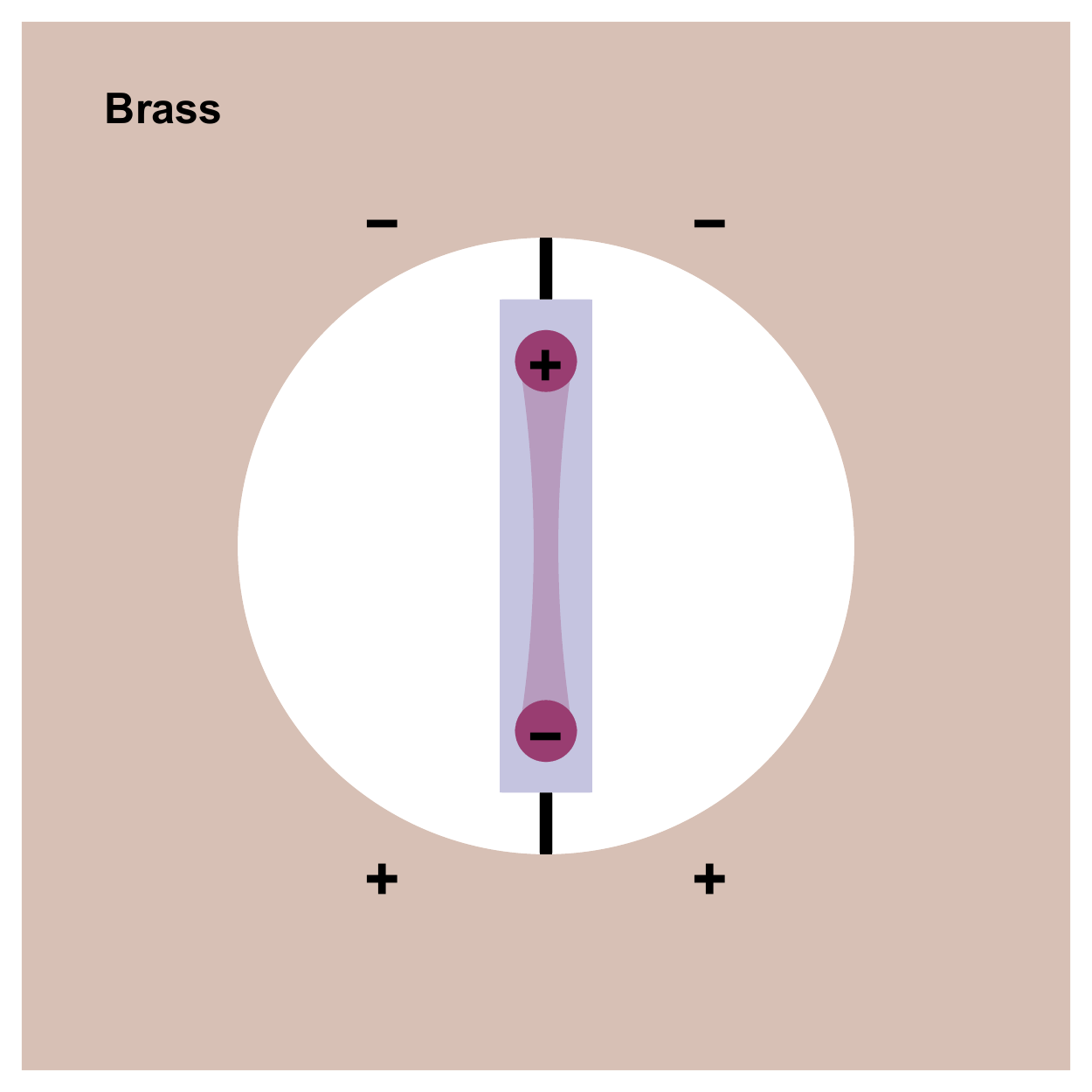} 
\end{center}
\caption{The quirk pair as an AC battery in brass (the CMS calorimeter material).  The oscillating charges draw electrons to one side, then back again.}  
\label{fig:Seaofbrass}
\end{figure}

A stopped quirk pair can also lose energy via interactions with the conductor that
surrounds it.  
We will consider energy loss due to the generation of an oscillating current 
within the material induced by the oscillating electromagnetic field generated by 
the quirks.  
The oscillating dipole field will give rise to a current in this conductor,
which in turn will give rise to energy loss \emph{via} Ohmic losses.
If we view the quirk pair as oscillating in a vacuum bubble of radius 1 \AA\ surrounded by 
a large volume of brass, see Fig.~\ref{fig:Seaofbrass}, we will see that the physics
of this system can be viewed as a driven RLC circuit. 
There are other mechanisms for energy loss
that we will not consider, 
{\it e.g.}~exciting phonons or ejecting bound electrons.

To get a rough estimate of the energy losses due to induced currents
in the conductor, we model the quirk as being in the center of a
conducting cavity with radius $a \sim 1$~\AA.
This cavity is intended as a crude model of the short-distance 
cutoff of the continuum description of the conductor due to the atomic
structure of the material.
The electric fields in the material will induce a surface charge
density on the boundary, which will screen the electric field
inside the conductor.
The quirks are non-relativistic, so we can describe the electric field
generated by them using a potential, and neglect magnetic fields.
We assume that the potential inside the cavity can be described by
the potential $\Phi = \Phi_d + \Phi_s$, where
\beq
\Phi_d(r,\theta,t) = \frac{1}{4\pi\epsilon_0} \frac{d(t)\cos\theta}{r^2},
\qquad
\Phi_s(r,\theta,t) = -\frac{1}{4\pi\epsilon_0} \frac{\bar{d}(t) r \cos\theta}{a^3},
\eeq
where $d = qR$ is the quirk dipole moment, and $\bar{d}$ is an ``image''
dipole moment that parameterizes the screening due to the electric charge
induced on the boundary of the conductor.
Here $\theta$ is the angle measured from the dipole moment direction.
To justify this, note that for $\bar{d} = d$ this potential describes the
potential of a static dipole in a conducting cavity with $\Phi = 0$ on the
boundary.
The electric potential inside the conductor is then determined by
\beq
\Phi(r, \theta, t) = \frac{1}{4\pi\epsilon_0} \frac{[d(t) - \bar{d}(t)]\cos\theta}{r^2}.
\eeq
The screening charge density is given by the discontinuity of the electric field
at the boundary:
\beq
\label{eq:rhos}
\rho_s = \epsilon_0 \, \Delta \vec{E} \cdot \hat{r}
= -\frac{3 \bar{d} \cos\theta}{4\pi a^3}.
\eeq
The screening charge is determined by the flow of current in the
conductor, which is in turn sensitive to the screening.
We are interested in the case where all quantities oscillate with
frequency $\omega$.
(The motion of the quirks is periodic but not sinusoidal, however, it is
an adequate approximation to estimate the energy loss for the dominant
frequency $\omega \sim 2\pi / T$, where $T$ is the period.)
The current inside the conductor is then
\beq
\vec{J} = \frac{\sigma}{1 - i \omega \tau_c} \vec{E},
\eeq
and we assume that the surface charge density $\rho_s$
on the boundary is given by
\beq
\label{eq:drhosdt}
\frac{\partial \rho_s}{\partial t} = -\vec{J} \cdot \hat{r}
= \frac{1}{4\pi\epsilon_0} \frac{\sigma}{1 - i \omega\tau_c}
\frac{2(d - \bar{d})\cos\theta}{a^3}.
\eeq
This is local conservation of charge under the assumption that there are
no surface currents.
Note that the $\theta$ dependence matches Eq.~(\ref{eq:rhos}), so this
model is consistent.
Eqs.~(\ref{eq:rhos}) and (\ref{eq:drhosdt}) then determine $\bar{d}$ in terms
of $d$:
\beq
-3i \omega \bar{d} = \frac{2\sigma/\epsilon_0}{1 - i \omega \tau_c} (d - \bar{d}).
\eeq
Using this we can compute the power dissipated by the current:
\beq
\label{eq:Pavg}
\langle P \rangle = \int dV\, \frac 12 {\text{Re}}(\vec{J} \cdot \vec{E}^*)
= \left( \frac{1}{4\pi\epsilon_0} \right)^2 \frac{4\pi\sigma (q R)^2}{3 a^3}
\frac{2}{1 + [\omega\tau_c - 2 \sigma / (3\omega\epsilon_0)]^2}.
\eeq
Note that this vanishes in both the limits $\omega \rightarrow 0$
and $\omega \rightarrow \infty$.
In the $\omega \rightarrow 0$ limit, the screening charge completely
shields the electric field inside the conductor;
in the $\omega \rightarrow \infty$ limit, the rapidly oscillating field
does not allow the drift electrons to move far, and the induced current
again vanishes.

The physics of this system can be understood as an RLC circuit.
The oscillating quirk dipole creates a potential difference on
the boundary of the conductor
\beq
V(t) = \frac{1}{4\pi\epsilon_0} \frac{qR}{a^2} \sin \omega t.
\eeq
This induces 
separated positive and negative charges
on the boundary, which can be viewed as a
capacitor with capacitance
\beq
C = 4\pi \epsilon_0 a.
\eeq
The capacitor is connected by the bulk of the conductor.
This has a resistance that can be approximated by assuming that it
is a shell of thickness $a$
\beq
R = \frac{3}{8\pi \sigma a}.
\eeq
where the numerical factors are chosen to get agreement with 
the model considered above.
At high frequencies, the damping in the current is equivalent to an
inductance
\beq
L = R \tau_c.
\eeq
The power dissipated in an RLC circuit is then given by
\beq
\langle P \rangle = \frac{ \frac 12 C^2 V_0^2 \omega^2 R}{1+(R^2C^2-2 LC) \omega^2  +C^2L^2 \omega^4} 
\eeq
which is identical to Eq.~(\ref{eq:Pavg}).

In brass, the power dissipated can be expressed numerically as
\beq
\vev{P(w)} \approx \frac{8.9 \times 10^{14} x^2  w^2 }{13000-220w^2 + w^4} \frac{\text{keV}}{s},
\eeq
where $w$ is the frequency of oscillation given in PHz ($10^{15} s^{-1})$, and $x$ is measured in \AA. The dipole oscillation frequency in the $\Lambda_{\text{IC}} \sim $ keV -- MeV range is in the PHz range,
\beq	
\omega \sim \frac{\pi}{T} \sim \frac{\Lambda_\text{IC}}{\sqrt{2 R m_Q}}
\sim 0.1~\text{PHz}
\left( \frac{\Lambda_\text{IC}}{\text{keV}} \right)
\left( \frac{m_Q}{\text{TeV}} \right)^{-1/2},
\eeq
so energy loss can be very efficient through current generation.  
Although this model has large uncertainties, we believe it is a reasonable
estimate of the order of magnitude of the energy loss due to induced
currents in the conductor.

\begin{figure}[t]
\begin{center}
\includegraphics[scale=1]{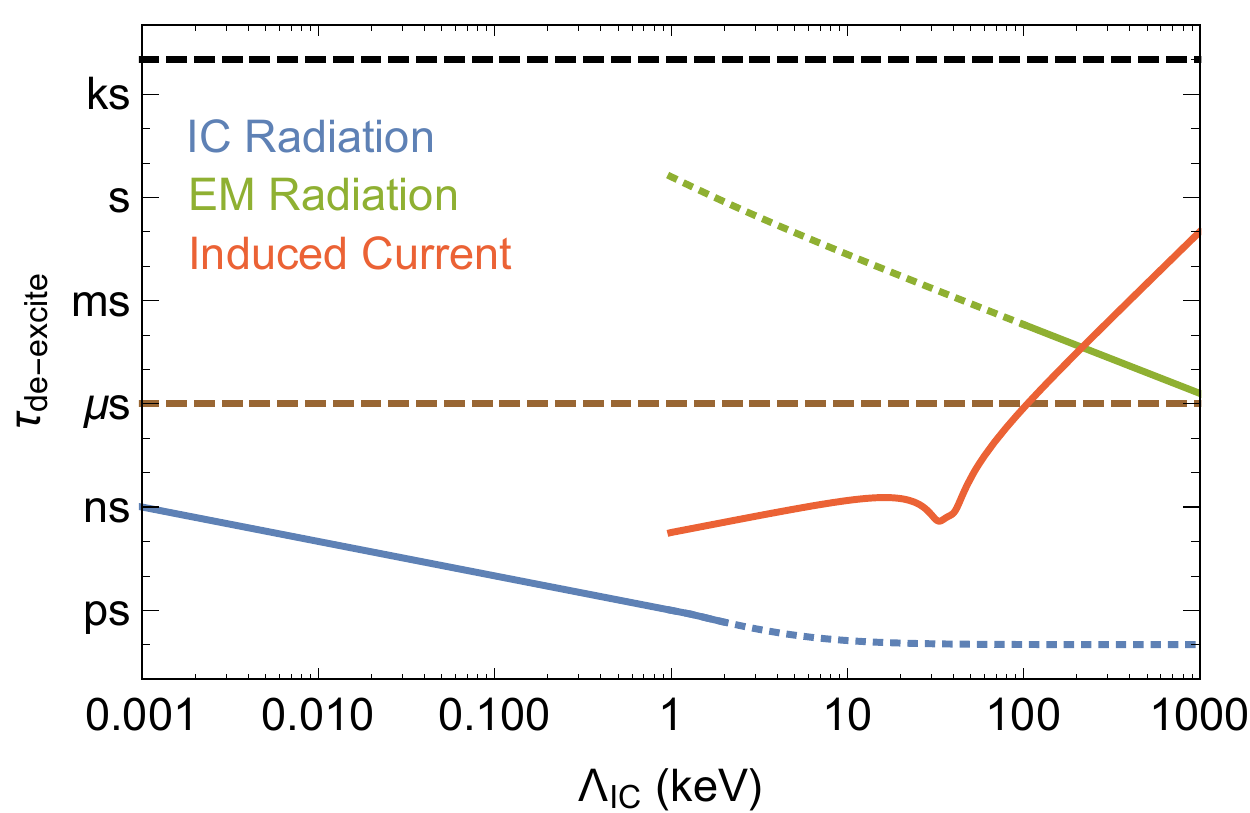} 
\end{center}
\caption{The different contributions to the de-excitation time through energy loss.  Below $\Lambda_\text{IC}\sim$~keV~$\sim$~\AA$^{-1}$, the infracolor glueball radiation dominates and leads to prompt annihilation.  Above 1 keV, the impact parameter becomes too large and infracolor glueball radiation ceases (indicated with the dotted line).  Similarly, below $\Lambda_\text{IC}\sim100$~keV, the conductor will screen propagating waves and our calculation may not be reliable.  Induced current losses dominate until $\gtrsim300$ keV when electromagnetic radiation losses become more important.  Importantly, across the entire range, $\tau_{\text{de-excite}}\ll 10^4 s$, where the stopped particle search begins to lose sensitivity due to very long particle lifetimes \cite{Sirunyan:2017sbs} (below dashed black line).  For $\Lambda_\text{IC}\gtrsim100$~keV, the stopped particle searches may become more sensitive than the results predicted here (above dashed brown line).}  
\label{fig:deexcitationtime}
\end{figure}

Using $dE =\Lambda_\text{IC}^2 dx$, the time to de-excite to the point where infracolor glueball emission is efficient is,
\beq
\tau \sim \int \frac{dE}{\vev{P(\omega)}}\sim \lp\frac{\Lambda_\text{IC}}{\text{keV}}\rp^3\frac{13000-220w^2 + w^4}{3.6 w^2} \times 10^{-15} s,
\eeq
and, if $\omega \gg 10^{16} s^{-1}$, 
\beq 
\tau \approx 3 \lp\frac{\Lambda_\text{IC}}{\text{keV}}\rp^5 \times 10^{-18} s.
\eeq   
In the most pessimistic case of $\Lambda_\text{IC}=\text{MeV}$, this results in a de-excitation time of $\tau_{\mbox{\scriptsize de-excite}} \sim 3$ ms, but is typically much lower.  For confinement scales $\Lambda_\text{IC}< 200$ keV, $\tau_{\mbox{\scriptsize de-excite}}\lesssim \mu$s, the timescale where de-excitation can be viewed as prompt.

\subsection{Summary}

The treatment of energy loss of a stopped quirk bound state is highly uncertain.  
The results of this appendix are summarized in Fig.~\ref{fig:deexcitationtime}.  
One important conclusion that is reliable is that the combination of 
effects discussed here will
ensure that the lifetime of such a bound state is shorter than $10^4$~s, so that 
the stopped particle search is efficient.
Infracolor glueball emission and interactions between the quirks and the material
may well give much more efficient energy loss mechanisms, but they are subject to
large uncertainties.
We therefore make the conservative assumption that the time for a stopped bound
state to de-excite is prompt.
We note that this may be significantly underestimating the reach, especially
for the larger values of $\Lambda_\text{IC}$ we consider.
A better understanding of these mechanisms would be very useful in interpreting
the results of this search, but we will not pursue this here.

  %%%%%%%%%%%%%%%%%%%%%%%%%%%%%%%%%%%%%%%

\bibliography{StoppingQuirks}

%%%%%%%%%%%%%%%%%%%%%%%%%%%%%%%%%%%%%%%

\end{document}